\documentclass[11pt]{article}
\usepackage{coling2016}
\usepackage{times}
\usepackage{url}
\usepackage{latexsym}
\usepackage{longtable}
\usepackage{verbatim}
\usepackage{natbib}
\usepackage{graphicx}

\title{Neural Information Retrieval: A Literature Review}

\author{\textnormal{Ye Zhang$^*$, Md Mustafizur Rahman, Alex Braylan,} \\ Brandon Dang, Heng-Lu Chang, Henna Kim, Quinten McNamara, Aaron Angert, \\ Edward Banner,  Vivek Khetan, Tyler McDonnell, An Thanh Nguyen, Dan Xu, \\ Byron C.\ Wallace, Matthew Lease$^\dagger$ \\
$^*$Department of Computer Science, University of Texas at Austin, yezhang@utexas.edu\\
$^\dagger$School of Information, University of Texas at Austin, ml@utexas.edu}
\providecommand{\keywords}[1]{\textbf{Keywords:} #1}

\date{}
\renewcommand*{\cite}{\citep}
\begin{document}
\maketitle
\newcommand{\wv}{{\tt word2vec}}
\newcommand{\pv}{{\tt ParagraphVector}}
\newcommand{\adhoc}{ad-hoc}
\newcommand{\Adhoc}{Ad-hoc}
\newcommand{\skipgram}{skip-gram}
\newcommand{\Skipgram}{Skip-gram}
\newcommand{\clueweb}{ClueWeb09-Cat-B}
\begin{abstract}
  A recent ``third wave'' of Neural Network (NN) approaches now delivers state-of-the-art performance in many machine learning tasks, spanning speech recognition, computer vision, and natural language processing. Because these modern NNs often comprise multiple interconnected layers, this new NN research is often referred to as {\em deep learning}. 
Stemming from this tide of NN work, a number of researchers have recently begun to investigate NN approaches to Information Retrieval (IR). While deep NNs have yet to achieve the same level of success in IR as seen in other areas, the recent surge of interest and work in NNs for IR suggest that this state of affairs may be quickly changing. In this work, we survey the current landscape of \emph{Neural IR} research, paying special attention to the use of learned representations of queries and documents (i.e., neural embeddings). We highlight the successes of neural IR thus far, catalog obstacles to its wider adoption, and suggest potentially promising directions for future research.
\end{abstract}

\section{Introduction}
\label{section:introduction}



We are in the midst of a tremendous resurgence of interest and renaissance in research on artificial neural network (NN) models for machine learning, now commonly referred to as {\em deep learning}\footnote{While not all NNs are `deep', and not all `deep' models are neural, the terms are often conflated in practice.}. While many valuable introductory readings, tutorials, and surveys already exist for deep learning at large, we are not familiar with any existing literature review surveying NN approaches to Information Retrieval (IR). Given the great recent rise of interest in such {\em Neural IR} from both researchers \citep{gao_deep_2015, li_deep_2016} and practitioners \citep{metz2016, ordentlich_network-efficient_2016} alike, we believe that such a literature review is now timely. IR researchers interested in getting started with Neural IR currently must identify and compile many scattered works. Unifying these into a coherent resource would provide a single point of reference for those interested in learning about these emerging approaches, as well as provide a valuable reference compendium for more experienced Neural IR researchers.
%

To address this need, this literature review surveys recent work in Neural IR. Our survey is intended for IR researchers 
already familiar with fundamental IR concepts and so requiring few definitions or explanations of these. Those less familiar with IR may wish to consult existing reference materials~\cite{croft2009search,manning2008} for unfamiliar terms or concepts. However, we anticipate many readers will have relatively less familiarity and experience with NNs and deep learning. Consequently, we briefly introduce key terms, definitions, and concepts in Sections \ref{section:word-embedding} and \ref{section:nn-background}. Because many exemplary introductory resources on deep learning already exist, we strive to avoid duplicating this material  and instead encourage readers to directly consult these resources for additional background. For general introductions to deep learning, see \citet{goodfellow_deep_2016,lecun_deep_2015,deng_tutorial_2014,yu_deep_2011,schmidhuber_deep_2015,arel_deep_2010}, and \citet{bengio_learning_2009}. See \citet{broder16-kdd-panel} for a recent panel discussion on deep learning. For introductions to deep learning approaches to other domains, see \citet{hinton_deep_2012} for automatic speech recognition (ASR), \citet{goldberg2015primer} for natural language processing (NLP), and \citet{wu2016google} for machine translation (MT). Regarding NN approaches to IR, informative talks and tutorials have been presented by \citet{gao_deep_2015,li_deep_2016}, and \citet{li_does_2016,li_opportunities_2016}. Many other useful tutorials and talks can be found online for general deep learning and specific domains. A variety of resources for deep learning, including links to popular open-source software, can be found online at \url{http://deeplearning.net}. 

In terms of scope, we restrict this survey to textual IR. For NN approaches to content-based image retrieval, see \citet{wan_deep_2014}. Similarly, we exclude work on NN approaches to acoustic or multi-modal IR, such as mixing text with imagery (see \citet{ma2015multimodal, Ma-learning-to-answer-2016}). We also intentionally focus on the current ``third wave'' revival of NN research, excluding earlier work. Finally, we largely follow the traditional divide between IR and NLP, including search-related IR research and excluding syntactic and semantic NLP work. However, this division is perhaps most difficult to enforce in regard to question answering (QA) and community QA (CQA) tasks, which perhaps represent the greatest cross-over between NLP and IR fields (e.g., see \citet{dumais2002web}). Recent years have seen QA research being more commonly published in NLP venues and often more focused on semantic understanding of short texts rather than search over large collections. As a pragmatic and imperfect compromise, we largely make the divide by publication venue, including those works appearing at IR venues and excluding those from NLP venues. For readers interested in further reading about deep QA, see \cite{bordes-chopra-weston:2014:EMNLP2014,kumar2015ask,gao_deep_2015,goldberg2015primer}.

Finally, regarding nomenclature, our use of {\em Neural IR}, referring to machine learning research on {\em artificial} NNs and deep learning, should not be confused with cognitive science research studying {\em actual} neural representations of relevance in the human brain (see \cite{Moshfeghi:2016:UIN:2911451.2911534}). In addition, note that the modern appellation of {\em deep learning} for the current ``third wave'' of NN research owes to these approaches using a large number of {\em hidden layers} in their NN architectures. For this literature review, we have chosen the term {\em neural} (rather than {\em deep}) because: i) most current work on textual NNs in NLP and IR is often actually quite shallow in the number of layers used (typically only a few hidden layers and often only one, though some notable recent exceptions exist, such as \citet{conneau2016very}); and ii) use of {\em neural} more clearly connects the current wave of {\em deep learning} to the long history of NN research. 


The remainder of this literature review is organized as follows. To illustrate the rough evolution of Neural IR research over time, we  begin by providing a concise history of Neural IR in Section \ref{section:history}. Following this, in Section~\ref{section:word-embedding} we introduce the concept of {\em word embeddings} and survey approaches that extend traditional IR models to integrate such embeddings, often using off-the-shelf \wv{} source code or trained embeddings. In contrast, Section~\ref{section:nn} surveys approaches which more fully embrace NN modeling. Such models treat word embeddings as the first layer in a network, to be learned along with all model parameters. This is an exciting direction, as researchers are broadly working toward developing {\em end-to-end} NN architectures for IR, which may depart significantly from traditional IR models. However, assuming a readership familiar with IR but possibly less versed in neural network (NN) concepts, Section \ref{section:nn-background} first briefly introduces several such concepts which underlie work in Section~\ref{section:nn}. Section~\ref{section:methods} provides discussion, and we conclude in Section~\ref{section:conclusion}. 
Table \ref{table:Data_Used} lists datasets used in Neural IR studies to date and the studies reporting on each. Tables \ref{table:code} and \ref{table:data}, respectively, list source code and data shared from published studies.

\section{A Brief History of Neural IR}
\label{section:history}



Introductory resources on deep learning cited in Section \ref{section:introduction} (see \citet{lecun_deep_2015} and \citet{goodfellow_deep_2016}) explain how the ``third wave'' of interest in neural network approaches arose. Key factors include increased availability of ``big data'', more powerful computing resources, and better NN models and parameter estimation techniques. While early use in language modeling for ASR and MT might be loosely related to language modeling for IR~\cite{ponte1998language}, state-of-the-art performance provided by neural language models trained on vast data could not be readily applied to the more typical sparse data setting of training document-specific language models in IR. Similarly, neural approaches in computer vision to learn higher-level representations (i.e., \emph{features}) from low-level pixels were not readily transferable to text-based research on words.

However, a rapid spread and adoption in NLP was sparked when \citet{mikolov2013distributed} and \citet{mikolov2013efficient} proposed \wv{}, a relatively simple model and estimation procedure for {\em word embeddings} (also known as {\em distributed term representations}), coupled with sharing both source code and trained embeddings (see Tables~\ref{table:code} and \ref{table:data}). This availability of \wv{} code and existing embeddings has provided one of the key avenues for early work on neural IR, especially for extending traditional IR models to integrate word embeddings. Given the significance and prevalence of such research, we review this body of work first in Section~\ref{section:word-embedding}. We also believe this organization is helpful to conceptually differentiate approaches that integrate word embeddings into traditional IR models vs.\ approaches directly incorporating word embeddings within NN models, reflecting a more significant shift toward pursuing {\em end-to-end} NN architectures in IR (Section~\ref{section:nn}).

\citet{salakhutdinov_semantic_2009} proposed the first ``third wave'' Neural IR work we are aware of, employing a deep {\em auto-encoder} architecture (Section~\ref{section:nn-background}) for semantic modeling for related document search (Section~\ref{section:nn-other-tasks}). They did not have relevance judgments for evaluation, so instead used document corpora with category labels and assumed relevance if a query and document had matching category labels. 

\Adhoc{} search was addressed more directly in 2013. \citet{huang_learning_2013} proposed a Deep Structured Semantic Model (DSSM)\footnote{\url{https://www.microsoft.com/en-us/research/project/dssm}}, which has generated a variety of follow-on work (e.g., ~\cite{gao_modeling_2014,shen_latent_2014,shen_learning_2014,mitra_exploring_2015,mitra_query_2015,song_multi-rate_2016}). DSSM's feed-forward NN learned low-dimensional vector representations (embeddings) of queries and documents, meant to capture the latent semantics in the texts. Because vocabulary size is typically quite large in real-world Web search, the authors were concerned that when using term vectors as input to the NN, the NN input layer size would become unmanageable for model training and inference. To address this, a ``word hashing'' dimensionality reduction technique was proposed to represent words based on their character trigrams (see Section~\ref{section:method-hashing}). Because such hashing also maps orthographic variants of the same word to nearby points in the hashing space, such hashing was also useful in handling out-of-vocabulary (OOV) query terms (how to best address OOV query terms not found in trained word embeddings remains an open question). While DSSM was applied to \adhoc{} websearch, the challenge of effectively supporting full-text search led the authors to instead index document titles only. The challenge to match \adhoc{} search effectiveness of traditional IR models with full document indexing using Neural IR approaches has been oft-discussed up to the present day (e.g., \ \citet{cohen_adaptability_2016,guo_deep_2016}).  
%

Also in 2013, \citet{clinchant_aggregating_2013} anticipated work on {\em word embedding} approaches to IR which would later stem from \wv{}. In this case, classic Latent Semantic Indexing (LSI)~\cite{deerwester1990indexing} was used to induce word embeddings, which were transformed into fixed-length Fisher Vectors (FVs) via \citet{jaakkola1999exploiting}'s Fisher Kernel (FK) framework, and then compared via cosine similarity for document ranking. 
Results for \adhoc{} search on three IR test collections were reported, though their proposed approach was outperformed by traditional Divergence From Randomness (DFR)~\cite{amati2002probabilistic} ranking. 
That same year, \citet{lu_deep_2013} also proposed {\em DeepMatch}. Similar to \citet{huang_learning_2013}, the authors focused on short-text matching, testing their approach on two datasets: community question answering (CQA) (matching questions with answers) and a Twitter-like micro-blog task (matching tweets with comments). 

In 2014, in addition to the DSSM follow-on works cited above \cite{shen_latent_2014,shen_learning_2014,gao_modeling_2014}, a variety of further work began to appear. \citet{sordoni_learning_2014} investigated query expansion, evaluating \adhoc{} search on TREC collections. \citet{le_distributed_2014} proposed their \pv{} method for composing word embeddings to induce semantic representations over longer textual units (see Section~\ref{section:method-representation}).  Their evaluation included an unusual IR task of detecting which of three results snippets did not belong for a given query (see {\em Outlier Detection} in Section~\ref{section:we-other-tasks}). \citet{li_deep_2014} used embeddings as input to a feed-forward deep NN to exploit a user's previous queries in order to improve document ranking. The first two neural IR papers also appeared at ACM SIGIR \cite{gupta_query_2014,zhang_continuous_2014}. \citet{gupta_query_2014} proposed an auto-encoder approach to mixed-script query expansion, considering transliterated search queries and learning a character-level ``topic'' joint distribution over features of both scripts. \citet{zhang_continuous_2014} considered the task of local text reuse, with three annotators labeling whether or not a given passage represents reuse. As in \citet{clinchant_aggregating_2013}, they adopted the Fisher Kernel (FK) 
to reduce variable-size word embedding concatenations to fixed length, and they further tested different hashing methods to reduce dimensionality. 

By 2015, work on neural IR had grown beyond what can be concisely described here. We saw \wv{} enter wider adoption in IR research (e.g., \citet{ganguly_word_2015,grbovic_search_2015,kenter_short_2015,zheng_learning_2015,zuccon_integrating_2015}), as well as a flourishing of neural IR work appearing at SIGIR \cite{ganguly_word_2015,grbovic_context-and_2015,mitra_exploring_2015,severyn_learning_2015,vulic_monolingual_2015,zheng_learning_2015}, spanning \adhoc{} search \cite{ganguly_word_2015,vulic_monolingual_2015,zheng_learning_2015,zuccon_integrating_2015}, QA sentence selection and Twitter reranking~\cite{kenter_short_2015}, cross-lingual IR \cite{vulic_monolingual_2015}, paraphrase detection \cite{kenter_short_2015}, query completion~\cite{mitra_exploring_2015}, query suggestion \cite{sordoni_hierarchical_2015}, and sponsored search \cite{grbovic_search_2015,grbovic_context-and_2015}. We also saw in 2015 the first workshop on Neural IR\footnote{\url{http://research.microsoft.com/en-us/um/beijing/events/DL-WSDM-2015/}}.

As of 2016, work on neural IR continues to accelerate in quantity of work, sophistication of methods, and practical effectiveness (e.g., \citet{guo_deep_2016}). SIGIR also featured its first workshop on the subject\footnote{\url{https://www.microsoft.com/en-us/research/event/neuir2016/}}, as well as a tutorial by  \citet{li_deep_2016}. To provide as current of coverage as possible in this literature review, we include articles appearing up through ACM ICTIR 2016 and CIKM 2016 conferences.

\section{Word Embedding Approaches to IR}
\label{section:word-embedding}

This section surveys recent studies extending traditional IR models to incorporate word embeddings. We distinguish these studies from approaches directly incorporating word embeddings within NN models (see Section~\ref{section:nn}), reflecting a more significant shift toward pursuing {\em end-to-end} NN architectures in IR. We begin by presenting background on word embeddings in Section~\ref{section:we-background}. Following this, we organize studies surveyed by IR tasks (see Table \ref{Tasks}). For additional discussion and comparison of methods, see Section~\ref{section:methods}.

\subsection{Background}
\label{section:we-background}

Terms have traditionally been encoded as discrete symbols which cannot be directly compared to one another (unless we consider edit distance at the character-level). Conceptually, this is equivalent to a {\em one-hot encoding} in which each term is represented as a sparse vector with dimensionality equal to the vocabulary size (each dimension corresponding to a unique word). To one-hot encode a given term $t$, we create a ${\bf 0}$ vector and set the index corresponding to $t$ to 1.  In such an encoding, all terms are orthogonal and equidistant to one another, and hence there is no easy way to recognize similar terms in the vector space. This has led to the infamous {\em vocabulary mismatch} problem in which an IR system must recognize when distinct but related terms occur in query and document to perform effective matching.

In contrast with such a one-hot representation, {\em word embeddings} (also known as {\em distributed term representations}) encode each symbol as a low-dimensional (say, hundreds of dimensions), continuous, dense vector. Following oft-cited \citet{firth1957synopsis}'s adage that ``You shall know a word by the company it keeps'', by capturing the extent to which words occur in similar contexts, word embeddings are able to encode semantic and syntactic similarity insofar as the embeddings for similar words will be nearby one another in vector space. This allows one to perform simple algebraic operations that reflect the word's meaning, e.g., \emph{wv}(``Madrid'') - \emph{wv}(``Spain'') + \emph{wv}(``France'') is expected to be close (e.g., with respect to Euclidean distance) to \emph{wv}(``Paris'').

Word embeddings are typically induced using large unlabeled corpora in an unsupervised way. The most popular method and software for learning word embeddings today, \wv{}, was proposed by \citet{mikolov2013distributed} and \citet{mikolov2013efficient}. \wv{} is a model that uses a three-layer NN with one hidden layer to learn word representations. Two variants were proposed: \emph{\skipgram{}} predicts the surrounding context given the current word, while \emph{continuous bag-of-words} (CBOW) predicts the current word given its context. The authors trained \wv{} on a large news corpus, releasing both software and learned embeddings online (see Tables~\ref{table:code} and \ref{table:data}).

As an alternative to \wv{}, \citet{pennington2014glove} proposed Global Vectors for Word Representation (GloVe), also sharing source code and learned embeddings online (see Tables~\ref{table:code} and \ref{table:data}). In contrast to \wv{}'s language modeling based on contextual windows, GloVe induces a log-bilinear regression model based on global word-word co-occurrence counts. The practical efficacy of using one embedding vs.\ another has been considered in different IR studies (see \citet{kenter_short_2015}), and it is not clear that any method or embedding set performs best in all situations. Models that jointly exploit multiple sets of embeddings may therefore be appealing \cite{zhang-roller-wallace:2016:N16-1,neelakantan-EtAl:2014:EMNLP2014}. See further discussion of word embeddings in Section~\ref{section:method-embedding}.


There is a long history of distributional approaches that is sometimes overlooked given current interest in \wv{}. As briefly mentioned in Section~\ref{section:history}, \citet{clinchant_aggregating_2013} use classic Latent Semantic Indexing (LSI)~\cite{deerwester1990indexing} to induce word embeddings, as well as \citet{jagarlamudi2011improving}'s Yule association measure. The authors also note \citet{collobert2008unified}'s embedding. In distributional semantics, see prior work by \citet{kiela2013detecting}, which \citet{lioma_non-compositional_2015} build upon in an IR context to infer whether query terms should be processed separately or as a single phrase for matching. In relation to prior contextual window approaches, such as the hyperspace analogue to language (HAL)~\cite{lund1996hyperspace} and probabilistic HAL \cite{azzopardi2005probabilistic} (see also \cite{bruza2002inferring}), \citet{zuccon_integrating_2015} provide a succinct but valuable comparison to \wv{}: co-occurrences in HAL within a window centered on a target term are accumulated, while \wv{} \skipgram{} fits the representation of a target term to predict its lexical context (and vice-versa for CBOW). \citet{zuccon_integrating_2015} conclude that as of 2015, ``it is not clear yet whether these neural inspired models are generally
better than traditional distributional semantic methods.'' 

%



{
\renewcommand{\arraystretch}{1.3}
\begin{table}[ht]
\centering
  \begin{tabular}{|p{3.5cm}| p{10.0cm} |}
    \hline    
    {\bf Task} & {\bf Studies} \\ \hline
	\Adhoc{} Retrieval & \citet{almasri_comparison_2016}, \citet{amer_toward_2016}, BWESG~(\citet{vulic_monolingual_2015}), \citet{clinchant_aggregating_2013}, \citet{diaz_query_2016}, GLM~(\citet{ganguly_word_2015}), \citet{mitra_dual_2016}, \citet{nalisnick_improving_2016}, NLTM~(\citet{zuccon_integrating_2015}),\citet{rekabsaz_uncertainty_2016}, \citet{roy_using_2016}, \citet{zamani_embedding-based_2016}, \citet{zamani_estimating_2016}, \citet{zheng_learning_2015}    \\
    \hline
    Bug Localization & \citet{ye_word_2016}\\
    \hline
    Contextual Suggestion & \citet{manotumruksa_modelling_2016}\\
    \hline
    Cross-lingual IR & BWESG~(\citet{vulic_monolingual_2015}) \\
    \hline
    Detecting Text Reuse & \citet{zhang_continuous_2014}\\
    \hline
    Domain-specific Semantic Similarity & \citet{de_vine_medical_2014}\\
    \hline
    Community Question Answering & \citet{zhou_learning_2015}\\
    \hline
    Short Text Similarity & \citet{kenter_short_2015}\\
    \hline
    Outlier Detection & \pv{}~(\citet{le_distributed_2014})\\
    \hline
    Sponsored Search & \citet{grbovic_context-and_2015}, \cite{grbovic_search_2015}\\
    \hline
  \end{tabular}
  \caption{IR tasks solved by embedding approaches. }
  \label{Tasks}
\end{table}
}

\subsection{\Adhoc{} Retrieval}
\label{section:we-task-adhoc}

\Adhoc{} retrieval refers to the initial search performed by a user: a single query, with no further interaction or feedback, on the basis of which an IR system strives to return an accurate document ranking. 

A series of international shared task evaluations for search systems have been established over the years, providing a wealth of reusable search collections, topics, and relevance judgments for training and evaluating IR systems (see \citet{sanderson2010test}):
\begin{itemize}
\item the Text REtreival Conference (TREC)\footnote{\url{http://trec.nist.gov}}~\cite{voorhees2005trec}, begun in 1992
\item the NII Testbeds and Community for Information access Research (NTCIR)\footnote{\url{http://research.nii.ac.jp/ntcir/index-en.html}}, begun in 1999
\item the Conference and Labs of the Evaluation Forum (CLEF), formerly known as the Cross-Language Evaluation Forum\footnote{\url{http://www.clef-initiative.eu}}~\cite{braschler2004cross}, begun in 2000
\item the Initiative for the Evaluation of XML Retrieval (INEX)\footnote{\url{http://inex.mmci.uni-saarland.de}}~\cite{lalmas2007evaluating}, begun in 2002, 
\item the Forum for Information Retrieval Evaluation (FIRE)\footnote{\url{http://fire.irsi.res.in}}, begun in 2008
\end{itemize}

A number of open-source toolkits have also been made available to enable and ease repeatable IR search experiments: Galago\footnote{\url{http://www.lemurproject.org/galago.php}}, 
Indri\footnote{\url{http://www.lemurproject.org/indri.php}}, 
Lemur\footnote{\url{http://www.lemurproject.org/lemur.php}}, 
Lucene\footnote{\url{http://lucene.apache.org}}, and
Terrier\footnote{\url{http://terrier.org}}.

As briefly mentioned in Section~\ref{section:history}, {\bf \citet{clinchant_aggregating_2013}} present recent pre-\wv{} work using word embeddings in IR. The authors articulate their primary contribution as showing that a document can be represented as a bag-of-embedded-words (BoEW). Latent Semantic Indexing (LSI)~\cite{deerwester1990indexing} is used to induce word embeddings, which are then transformed into fixed-length Fisher Vectors (FVs) via \citet{jaakkola1999exploiting}'s Fisher Kernel (FK) framework. 
The FVs are then compared via cosine similarity for document ranking. Experiments on \adhoc{} search using Lemur are reported for three collections: TREC ROBUST04, TREC Disks 1\&2, and English CLEF 2003 \Adhoc. While results show improvement over standard LSI on all three collections, standard TF-IDF performs slightly better on two of the three collections, and Divergence From Randomness (DFR)~\cite{amati2002probabilistic} performs far better on all collections. The authors note, ``These results are not surprising as it has been shown experimentally in many studies that latent-based approaches such as LSI are generally outperformed by state-of-the-art IR models in Ad-Hoc tasks.'' Document clustering results are also reported. The authors note that their approach is independent of any particular embedding technique, and briefly report on using a word embedding based on \citet{jagarlamudi2011improving}'s Yule association measure instead of LSI. The authors propose to consider further embeddings in future work, such as \citet{collobert2008unified}'s embedding. Finally, the authors note that another potential advantage of their proposed framework is the ability to seamlessly deal with multilingual documents (e.g., see \citet{vulic_monolingual_2015}).


{\bf \citet{ganguly_word_2015}} propose a Generalized Language Model (GLM) for integrating word embeddings with query-likelihood language modeling. Semantic similarity between query and document/collection terms is measured by cosine similarity between word embeddings induced via \wv{} CBOW. The authors frame their approach in the context of classic {\em global} vs. {\em local} term similarity, with word embeddings trained without reference to queries representing a global approach akin to the Latent Dirichlet Allocation (LDA) of \citet{wei2006lda}. Like \citet{rekabsaz_uncertainty_2016} and \citet{zuccon_integrating_2015}, the authors build on \citet{berger1999information}'s ``noisy channel'' translation model. The noisy channel may transform a term $t'$ observed in a document into a term $t$ observed in a query, either by document sampling or collection sampling. For document sampling, they take $P(t,t'|d) = P(t|t',d)P(t'|d)$, where $P (t|t',d)$ is computed based on cosine similarity between embeddings of $t$ and $t'$. 
For a single document, $P(t, t'|d)$  considers all document terms, but for transformation via the collection $C$, $P(t, t'|C)$ considers only a small neighborhood of terms around query term $t$ to reduce computation. 
Smoothing of mixture model components resembles classic cluster-based LM smoothing of \citet{liu2004cluster}. It is unclear how out-of-vocabulary (OOV) query terms are handled. \Adhoc{} search results reported for TREC 6-8 and Robust using Lucene show improvement over both unigram query-likelihood and LDA. However, parameters appear to be tuned on testing collections, and LDA results are much lower than \citet{wei2006lda}'s, which the authors hypothesize is due to training LDA on the entire collection rather than collection subsets. The authors do not compare their global approach vs.\ local pseudo-relevance feedback (PRF)~\cite{lavrenko2001relevance}, which prior has shown to outperform LDA \cite{yi2009comparative}, and while typically a query-time technique, can be approximated for greater efficiency \cite{cartright2010fast}. 

{\bf \citet{zuccon_integrating_2015}} propose a Neural Language Translation Model (NLTM), also integrating word embeddings into \citet{berger1999information}'s classic translation model approach to query-likelihood IR. 
They estimate the translation probability between terms as the cosine similarity of the two terms divided by the sum of the cosine similarities between the translating term and all of the terms in the vocabulary. However, note that use of cosine similarity as a proxy for a probability assumes word embeddings have purely positive values. Previous state-of-the-art translation models use mutual information (MI) embeddings to estimate translation probability. Experiments evaluating the NLTM approach on \adhoc{} search are reported on the TREC datasets AP87-88, WSJ87-92, DOTGOV, and MedTrack. Results indicate that the NLTM approach provides moderate improvements over the MI and classic TM systems, based on modest improvements to a large number of topics, rather than large differences on a few topics. Sensitivity analysis of the various model hyper-parameters for inducing word embeddings shows that manipulations of embedding dimensionality, context window size, and model objective (CBOW vs \skipgram{}) have no consistent impact upon NLTM's performance vs.\ baselines. Regarding choice of training corpus for learning embeddings vs.\ search effectiveness, although effectiveness typically appears highest when embeddings are estimated using the same collection in which search is to be performed, the differences are not statistically significant. Source code and learned embeddings are shared online (see Tables~\ref{table:code} and \ref{table:data}). 

Similar to {\em RegressionRank} \cite{lease2009regression}, {\bf \citet{zheng_learning_2015}} propose DeepTR to learn effective query term weights through supervision. However, whereas \citet{lease2009regression} engineer features, \citet{zheng_learning_2015} automatically construct feature vectors using \wv{} CBOW word embeddings. In particular, each query term's feature vector is constructed by simple subtraction of the average embedding vectors of query terms from the given term's embedding vector. L1 LASSO regression is used to predict a target weight for each query term given its feature vector. The {\em recall weight}~\cite{zhao2010term} of each term, 
learned from relevance judgments, is used as the target term weight for regression. In-collection training of target weights is assumed (i.e., it is assumed that relevance judgments are available for the same collection which is to be searched), using 5-fold cross-validation, with 80\% of queries (and their relevance judgments) used for training and 20\% used for testing.
%
%
%
Experiments on \adhoc{} search are performed with Indri on 4 TREC test collections: Robust04, WT10t, GOV2, and \clueweb{}, comparing against 3 baselines: BM25, unigram language modeling (LM), and \citet{metzler2005markov}'s sequential dependency model. Results are reported only for verbose TREC {\tt Description} queries, rather than keyword {\tt title} queries (and analysis by the authors shows that DeepTR performs much better on queries of 4 or more words). DeepTR is shown to achieve statistically significant improvements over baselines, showing that term embeddings can be effectively exploited to improve \adhoc{} search accuracy without requiring end-to-end neural network modeling. Regarding embedding dimensionality, \citet{zheng_learning_2015} find that 100 dimensions work best for estimating term weights, better than 300 and 500. Training on external corpora performs best, though no single corpus is consistently best, despite widely varying size of training data. No psuedo-relevance feedback~\cite{lavrenko2001relevance} experiments are reported, though the authors suggest that embeddings may help identify expansion terms. 

A pair of studies by {\bf Zamani and Croft~\cite{zamani_embedding-based_2016,zamani_estimating_2016}} investigate related word embedding approaches. %
Similar to \citet{zheng_learning_2015}, {\bf \citet{zamani_estimating_2016}} develop a method for query embedding based on the embedding vectors of its individual words. However, whereas \citet{zheng_learning_2015}'s simple average of vectors is heuristic, \citet{zamani_estimating_2016} develop a theoretical framework. Their intuitive idea is that a good query vector should yield a probability distribution $P(w|q) = \delta(w,q) / Z$ induced over the vocabulary terms similar to the query language model probability distribution (as measured by the KL Divergence). Here, $w$ is the embedding vector of the word, $q$ is the query embedding vector, $\delta(w,q)$ is the similarity between the two embedding vectors, and $Z$ is a normalization constant. 
%
For similarity, they use the softmax and sigmoid transformations of cosine similarity. The authors show that the common heuristic of averaging individual query term vectors to induce the overall query embedding is actually a special case of their proposed theoretical framework for the case when vector similarity is measured by softmax and the query language model is estimated by maximum likelihood. Experiments with \adhoc{} search using Galago are reported for three TREC test collections: AP, Robust04, and GOV2, using keyword TREC {\tt title} queries. Word embeddings are induced by GloVe~\cite{pennington2014glove}. Two different methods are used for estimating the query language model: via Pseudo Relevance Feedback (PRF)~\cite{lavrenko2001relevance}, which they refer to as {\em pseudo query vector} (PQV), and via maximum-likelihood estimation (MLE). Two different methods are used for calculating similarity: Softmax and Sigmoid. Softmax is easy to use because of the closed form solution, but the sigmoid function consistently showed better performance, likely due to the flexibility provided by its two free parameters. Although PQV  shows higher mean average precision, precision of top documents appears to suffer. Results also emphasize the importance of training data selection. Embeddings trained from the same domain as the documents being searched perform better than embeddings trained on a larger dataset from a different domain. Increasing the dimensionality of embedding vectors seems to improve retrieval, but it is unclear what would happen beyond the 300 dimensions considered. The authors also present query classification results (see Section~\ref{section:we-other-tasks}).
With regard to future work, because the current work assumes that embedding vectors for all query terms are available, one future direction could explore smoothing instead. The normalization term $Z$ is also ignored by making a simplifying assumption that $Z$ is the same across all queries, which does not hold in practice. Future research might seek to compute $Z$ more efficiently or propose an alternative embedding distribution. 

%
In their other study, {\bf \citet{zamani_embedding-based_2016}} characterize their primary contribution vs.\ prior work as being their focus on applying embedding to queries rather than documents, estimating query language models which efficiently outperform embedding-based document language models~\cite{ganguly_word_2015}. Moreover, whereas \citet{almasri_comparison_2016}  propose a heuristic method for query expansion VEXP based on similarity of word embeddings, \citet{zamani_embedding-based_2016} consider expansion of the whole query rather than term-by-term expansion. Also notable, while similarity of embedding vectors is often measured by cosine or Euclidean distance, the authors propose transforming the similarity values using sigmoid and show the empirical benefit of doing so. Two embedding-based query expansion (EQE) approaches are proposed. Similar to RM1 and RM2 in \citet{lavrenko2001relevance}'s relevance models, \citet{zamani_embedding-based_2016} EQE1 assumes conditional independence between query terms, whereas EQE2 assumes the semantic similarity between two terms is independent of the query. The authors also propose an embedding-based relevance model (ERM). As in \citet{zamani_estimating_2016}, \adhoc{} search experiments with Galago are reported for TREC AP, Robust04, and GOV2 test collections using keyword TREC {\tt title} queries, with word embeddings induced via GloVe~\cite{pennington2014glove}. Baselines include \citet{ganguly_word_2015}'s GLM and \citet{almasri_comparison_2016}'s VEXP. The authors also consider an unsupervised variant baseline of \citet{zheng_learning_2015} based on the similarity of vocabulary term vectors and the average embedding vector of all query terms (AWE). PRF experiments compare RM1 and RM2 vs.\ ERM. Proposed methods tend to outperform baselines, and EQE1 tends to outperform EQE2. While the authors consider training GloVe on three external corpora, they find that ``there is no significant differences between the values obtained by employing different corpora for learning the
embedding vectors.'' The authors also present a sensitivity analysis of the sigmoid parameters. For future work, the authors propose modifying the learning process of embedding vectors to yield discriminative similarity values suited to many IR tasks. They also suggest further theoretical analysis of their use of sigmoid.

{\bf \citet{nalisnick_improving_2016,mitra_dual_2016}} propose a Dual Embedding Space Model (DESM), writing that ``a crucial detail often overlooked when using \wv{} is that there are two different sets of vectors... {\sc in} and {\sc out} embedding spaces [produced by \wv{}]... By default, \wv{} discards $W_{OUT}$ at the end of training and outputs only $W_{IN}$...'' In contrast, the authors retain both input and output embeddings. Query terms are mapped to the input space and document words to the output space. Documents are embedded by taking an average (weighted by document term frequency) over embedding vectors of document terms. Query-document relevance is computed by average cosine similarity between each query term and the document embedding. The authors induce word embeddings via \wv{} CBOW only, though note that \skipgram{} embeddings could be used interchangeably. Embeddings learned on a Bing query log are compared to embeddings learned on a proprietary Web corpus. Results show ``DESM to be a poor standalone ranking signal on a larger set of documents'', so a re-ranking approach is proposed in which the collection is first pruned to all documents retrieved by an initial Bing search, and then re-ranked by DESM. Experiments with \adhoc{} search are carried out on a proprietary Web collection using both explicit and implicit relevance judgments. In contrast with \cite{huang_learning_2013}'s DSSM, full Webpage texts are indexed instead of only page titles. DESM's {\sc in} and {\sc out} embedding space combinations are compared to baseline retrieval by BM25 and latent semantic analysis (LSA)~\cite{deerwester1990indexing}. Out-of-vocabulary (OOV) query terms are ignored for their DESM approach but retained for baselines. Re-ranking results show improvement over baselines, especially on the implicit feedback test set, with best performance when word embeddings are trained on queries and using {\sc in-out} embedding spaces in document ranking. The authors surmise that query-training performs better due to users tending to include only significant terms from their queries. For future work, the authors propose investigating using of {\sc in} and {\sc out} embeddings in pseudo-relevance feedback and query expansion. They also propose investigating further ways to induce document embeddings and to seek a principled way to avoid the need for two-stage re-ranking. Learned word embeddings are shared online (see Table~\ref{table:data}).

{\bf \citet{vulic_monolingual_2015}} present a unified framework for monolingual and cross-lingual IR based on word embeddings. Section~\ref{section:we-cross-lingual} discusses their approach to learning bilingual embeddings and their findings for cross-lingual search. We focus here on their monolingual model and results. Word-embeddings are learned via \wv{}'s \skipgram{} model (the approach does not appear to be specific to \skipgram{}, though use of CBOW or GloVe embeddings is not discussed). Given term embeddings, query and document embeddings are induced by simply adding their constituent term vectors (we assume query and document vectors are normalized to avoid length bias, though this is not discussed). They also propose an extended approach in which the sum is weighted by each term's information-theoretic self-information, which akin to IDF is expected to signify more important terms. Documents are then ranked by cosine similarity between query and document embedding vectors. Experiments on \adhoc{} search are performed on CLEF 2001-2003 test collections, reporting unigram language modeling and LDA~\cite{wei2006lda} as baselines. The authors also report mixture models combining baselines and combining unigram modeling with their embedding approach, tuning the mixture weight. Queries used appear to be the concatenation of both {\tt title} and {\tt description} topic fields, so verbose but with key terms repeated. \wv{} dimensions are varied from 100-800 in steps of 100, and window size is varied from 10-100 in steps of 10. Parameters appear to be tuned on testing data, suggesting upperbound performance achievable by each method (rather than expected performance in practice). The proposed embedding approach outperforms LDA, is outperformed by unigram modeling, and performs best in mixture with the unigram. Composing vectors weighted by self-information delivers small but consistent improvements over simple addition. For future work, the authors propose to investigate other vector composition models and use of embeddings in pseudo-relevance feedback.

{\bf \citet{diaz_query_2016}} learn query-specific {\em local} word embeddings, rather than the typical practice of inducing {\em global} embeddings from a corpus without reference to user queries (see useful related discussion on local vs.\ global term similarity in \cite{ganguly_representing_2016,zuccon_integrating_2015}). This is the only work to date we are aware of that learns and exploits topic-specific word embeddings for \adhoc{} search. The authors use the document retrieval scores from each query to learn a document relevance probability distribution over the document collection. They then sample a set of $k$ documents according to this probability distribution, training word embeddings on this topically-biased document subset. The locally trained words embeddings are then used to derive an expansion language model for the query, which is interpolated with the original query language model in usual fashion to mitigate risk of {\em query drift}. Reported experiments use three TREC datasets: TREC1\&2, Robust, and \clueweb{}. The authors train local word embedding on these three test collections, as well as on two external corpora: Gigaword and Wikipedia. Similar to \citet{singhal1999document}, queries are run first on external documents to retrieve related documents for expansion, which are then used to improve the expansion of queries to be run on the test collection of interest. 
%
The proposed query expansion strategy outperforms a {\em global} baseline in which word embeddings are induced irrespective of queries. While these results are certainly encouraging, the computational cost of inducing word embeddings at query-time is expensive. Consequently, the authors suggest future work should address efficiency, perhaps by pre-computing some smaller number of topical embeddings at coarser granularity, and then simply selecting between these alternative topical embeddings at query-time.

As mentioned in discussion of \citet{zamani_embedding-based_2016}, {\bf \citet{almasri_comparison_2016}} propose a heuristic method VEXP for term-by-term query expansion using embedding vectors. For each query term, it collects its several most similar terms in the embedding space and adds them to the query. Experiments with \adhoc{} search use Indri on four CLEF medical test collections: Image2010-2012 (short documents and queries, text-only) and Case2011 (long documents and queries). Baselines include pseudo-relevance feedback~\cite{lavrenko2001relevance} and mutual information. They evaluate both CBOW and \skipgram{} \wv{} embeddings (using default dimensionality and context window settings) but present only \skipgram{} results, noting ``there was no big difference in retrieval performance between the two''. The authors consider adding a fixed number of 1-10 expansion terms per query term and also compare two smoothing methods: linear Jelineck-Mercer vs.\ Dirichlet. Results show VEXP achieves higher Mean Average Precision (MAP) than other methods.

%
{\bf \citet{roy_using_2016}} propose three approaches to utilize word embeddings in query expansion based on $K$-Nearest Neighbor (KNN). The first approach (i) computes the $K$ nearest neighbors for each query term in the word embedding space. For each nearest neighbor, they calculate the average cosine similarity vs.\ all query terms, selecting the top-$K$ terms according to average cosine score. The second approach (ii) reduces the vocabulary space considered by KNN by only considering terms appearing in the top $M$ pseudo-relevant documents retrieved by the query. In the third approach (iii), an iterative (computationally expensive) pruning strategy is applied to reduce the number of nearest neighbors, assuming that nearest neighbors are similar to one another. Search is performed using unigram language model retrieval with Jelinek-Mercer smoothing. Baselines include no-expansion unigram and RM3 interpolated query expansion \cite{abdul2004umass} between unigram and RM1 relevance model~\cite{lavrenko2001relevance}. Negative results show that the standard RM3 model performs better, suggesting that word embeddings do not yield improvements in this formulation.

{\bf \citet{amer_toward_2016}} investigate word embedding for personalized query expansion in the domain of social book search\footnote{\url{http://social-book-search.humanities.uva.nl}}. While personalized query expansion is not new \cite{chirita2007personalized, carmel2009personalized}, use of word embedding for personalization is novel. The proposed method consists of three steps: user modeling, term filtering and selection of expansion terms. A user is modeled as a collection of documents, and query terms are filtered to remove adjectives, which may lead to noisy expansion terms. For each remaining query term, similar to \citet{roy_using_2016}'s KNN approach, the top-$K$ most similar terms are selected based on cosine similarity in the embedding space. Evaluation on the social book search task compares \wv{} trained on personalized vs.\ non-personalized training sets. However, results show that expansion via word embeddings strictly hurts performance vs.\ no expansion at all, in contrast with findings of \citet{roy_using_2016} and \citet{mitra_query_2015}. This may stem from training \wv{} embeddings only on social book search documents. Results further suggest that personalized query expansion does not provide improvement over non-personalized query expansion using word embedding. The authors postulate that sparse training data for personalization is the main problem here and leave this for future work.

{\bf \citet{rekabsaz_uncertainty_2016}} recommend choosing similar terms based on a global similarity threshold rather than by KNN because some terms should naturally have more similar terms than others. They choose the threshold by setting it such that for any term, the expected number of related terms within the threshold is equal to the average number of synonyms over all words in the language. This method avoids having to constrain or prune the KNN technique as in \cite{roy_using_2016}. They use multiple initializations of the \wv{} \skipgram{} model to produce a probability distribution used to calculate the expected cosine similarity, making the measure more robust against noise. Experiments on TREC 6-8 and HARD 2005 incorporate this threshold-setting method into a translation language model \cite{berger1999information} for \adhoc{} retrieval and compare against both a language model baseline and a translation language model that uses KNN to select similar words. The threshold-based translation language model achieves the highest Mean Average Precision (MAP).


\subsection{Sponsored Search} 

\newcommand{\qv}{{\tt query2vec}}
\newcommand{\aqv}{{\tt ad-query2vec}}
\newcommand{\daqv}{{\tt directed ad-query2vec}}
\newcommand{\contexttovec}{{\tt context2vec}}
\newcommand{\contenttovec}{{\tt content2vec}}
\newcommand{\ccv}{{\tt context-content2vec}}

Grbovic et al. present a pair of studies in sponsored search \cite{grbovic_search_2015,grbovic_context-and_2015}. In 
their first study, {\bf \citet{grbovic_search_2015}} propose \qv{}, a two-layer architecture for search retargeting, where the upper layer models the temporal context of a query session using a \wv{} \skipgram model, and the lower layer models word sequences within a query using \wv{} CBOW. They also introduce two incremental models: \aqv{}, which incorporates the learning of ad click vectors in the upper layer by inserting them into query sequences after queries that occurred immediately prior to an ad click; and \daqv{}, which uses past queries as context for a directed language model in the upper layer. The models are trained using 12 billion sessions collected on Yahoo search and evaluated offline using historical activity logs, where success is measured by the click-through rate of ads served. All three \qv{} models show improvement over sponsored keyword lists and search retargeting using \wv{} and query flow graph.

In their subsequent, longer study, {\bf \citet{grbovic_context-and_2015}} propose a method to train context and content-aware word embeddings. The first model they propose is \contexttovec{}. It treats a search session as a sentence and each query from the session as a word from the sentence. It uses \wv{}'s \skipgram{} model. Queries with similar context will result in similar embeddings. The second model is \contenttovec{}. This method is similar to \citet{le_distributed_2014}'s \pv{} in that it uses the query as a paragraph to predict the word in its content. The third model \ccv{}, similar to their earlier \qv{}, combines \contexttovec{} and \contenttovec{} to build a two-layer model which jointly considers the query session context and the query context. To generate query rewrites, they use $K$-Nearest Neighbors (KNN) in the embedding space. They train embeddings on query content and search session data. They also incorporate ad click and search link click events as additional context to improve the query representation for a specific task. After finishing training embeddings, they evaluate them on an in-house data set and TREC Web Track dataset from 2009-2013. The best performance is achieved by \ccv{} with ad clicks and search link clicks in terms of NDCG and editorial grade. They also show that by considering ad clicks, the query rewrites generated can cover more bid terms.

\subsection{Other Tasks} 
\label{section:we-other-tasks}


\textbf{Community Question Answering (CQA).} Learning of word embeddings coupled with category metadata for CQA is proposed by {\bf \citet{zhou_learning_2015}}. They adopt \wv{}'s \skipgram{} model augmented with category metadata from online questions, with category information encoding the attributes of words in the question (see \citet{zhang_deep_2016} for another example of integrating categorical data with word embeddings). In this way, they group similar words based on their categories. They incorporate the category constraint into the original \skipgram{} objective function. After the word embedding is learned, they use Fisher kernel (FK) framework to convert the question into a fixed length vector (similar to \citet{clinchant_aggregating_2013} and \citet{zhang_continuous_2014}). To retrieve similar questions, they use the dot product of FVs to calculate the semantic similarities. For the experiment, they train the word embeddings on Yahoo! Answers and Baidu Zhidao for English and Chinese respectively. 
Results show that the category metadata powered model outperforms all the other baselines not using metadata. Future work might include exploring how to utilize other metadata information, such as user ratings, to train more powerful word embeddings.

\textbf{Contextual Suggestion.}
For the task of context-aware venue recommendation, users can express a set of contextual aspects for their preferences, where each aspect has multiple contextual dimension terms, and each term has a list of related terms. The goal is to rank a list of venues by measuring how well each venue matches the user's contextual preferences. The traditional approach for this task, collaborative filtering, suffers from data sparsity. 
{\bf \citet{manotumruksa_modelling_2016}} model user preferences using word embeddings, the only work we are aware of utilizing word embeddings for this task.
They develop two approaches to model user-venue preferences and context-venue preferences using word embeddings.
First, they infer a vector representation for each venue from user comments on it and model the user-venue preferences using the rated venues' vector representation in the user's profile.
Second, they model the dimensions of each aspect in the context-venue preferences by identifying top similar terms for that dimension.
To evaluate the user-venue preferences model, they first train word embeddings using \wv{} \skipgram{} on Foursquare data giving venue comments. Then they calculate the cosine similarity between the vector representation of venue and the user and use it as a feature in the learning to rank system. They conduct experiments using Terrier on TREC 2013 and 2014 Contextual Suggestion tracks\footnote{\url{https://sites.google.com/site/treccontext/}}. 
Results show that the system using word embeddings outperforms those without using embeddings.
To evaluate the context-aware preference model, the authors use cosine similarity between the venue and each of the contextual aspect as a feature in the ranking system. They also incorporate venue-dependent features and user-venue preference features. Results on TREC 2015 Contextual Suggestion task show that the proposed new system outperforms the baseline which does not utilize user information and contextual preferences.

\textbf{Cross-Lingual IR.}
\label{section:we-cross-lingual}
A unified framework for monolingual and cross-lingual IR using word embeddings is proposed by {\bf \citet{vulic_monolingual_2015}}. The monolingual model and results are presented in Section~\ref{section:we-task-adhoc}. Here, we discuss the authors' approach to learning bilingual embeddings and their findings for cross-lingual search. Their Bilingual Word Embeddings \Skipgram{} (BWESG) approach departs from prior work requiring sentence-aligned, parallel bilingual corpora (or bilingual dictionaries) to learn bilingual embeddings. Instead, only document-aligned {\em comparable} bilingual documents are needed. Their key idea is a \textit{merge and shuffle} process in which matched documents from each language are first merged and sentence boundaries removed. Next, they randomly shuffle the words in the constructed pseudo-bilingual document. This assures that each word, regardless of language, obtains word {\em collocates} from both languages. While this approach is simple and able to benefit from a potentially vaster body of comparable vs.\ parallel corpora for training, random shuffling loses the precise local context windows exploited by \wv{} training (which parallel corpora would provide), effectively setting context window size to the length of the entire document. The approach and experimental setup otherwise follows the monolingual version of the authors' method. Cross-lingual results for CLEF 2001-2003 \Adhoc{} English-Dutch show that the embedding approach outperforms the unigram baseline and is comparable to the LDA baseline. As with monolingual results, the mixture models perform better, with a cross-lingual three way mixture of unigram, LDA, and embedding. No experiments are reported with parallel corpora for comparison, which would be interesting for future work. 


\textbf{Detecting Text Reuse.} 
The goal of {\bf \citet{zhang_continuous_2014}} is to efficiently retrieve passages that are semantically similar to a query, making use of hashing methods on word vectors that are learned in advance. Other than the given word vectors, no further deep learning is used. As in \citet{clinchant_aggregating_2013} and \citet{zhou_learning_2015}, they adopt the Fisher Kernel framework 
to convert variable-size concatenations of word embeddings to fixed length. However, this resulting fixed-length Fisher vector is very high-dimensional and dense, so they test various state-of-the-art hashing methods (e.g., Spectral Hashing \cite{weiss2009spectral} and SimHash \cite{charikar2002similarity}) for reducing the Fisher vector to a lower-dimensional binary vector. Experiments are conducted on six collections including TIPSTER
(Volumes 1-3), \clueweb{}, Tweets2011, SogouT 2.0,
Baidu Zhidao, and Sina Weibo, with some sentences manually annotated for semantic similarity. Hashing methods that use Fisher vector representations based on word embeddings achieve higher precision-recall curves than hashing methods without vector representations and have comparable computational efficiency.

\textbf{Domain-specific semantic similarity.}
Word embeddings have also shown promise in capturing semantic similarities in specific domains of information retrieval. {\bf \citet{de_vine_medical_2014}} illustrate the ability of the \wv{} \skipgram{} embedding to learn medical embeddings from patient records and medical journal abstracts. The embeddings produce semantic similarities that strongly correlate with medical expert evaluations. Specifically, they correlate better with expert-provided concept similarities than previous corpus-driven methods for semantic similarity such as latent semantic analysis (LSA). To achieve this, prior to training semantic embeddings, documents are preprocessed through a medical concept tagger, such that documents are transformed into sequences of medical concept identifiers. Thus, the authors also demonstrate the potential for word embeddings to be trained from structured ontologies rather than raw text. Performance differences between these approaches is left for future work. Additional future work includes evaluation of the effect of medical embeddings on the performance of medical IR systems.


\textbf{Outlier Detection.} The evaluation of {\bf \citet{le_distributed_2014}}'s proposed \pv{} (PV) model includes an unusual outlier detection IR task. Section~\ref{section:method-representation} describes the PV model. In the task evaluation, each instance is a triplet of search result snippets, with two snippets deriving from the same query, and the third coming from a different query. The goal of the task is to detect the outlier snippet from each triplet. Results show that PV outperforms bag-of-words and bigram models.  

\textbf{Query classification.} Using data from the KDD Cup 2005 Task\footnote{\url{http://www.kdd.org/kdd-cup/view/kdd-cup-2005}}, {\bf \citet{zamani_estimating_2016}} propose an embedding method for categorizing queries. See Section~\ref{section:we-task-adhoc} for description of \citet{zamani_estimating_2016}'s overall method and results for \adhoc{} search. For query classification, given a training data pair $\langle query,category \rangle$, the authors first calculate the centroid vector of all query embedding vectors under a category. Then for a test query, they use the query embedding form and calculate the distance to the $K$ nearest neighbor centroid vector. Finally, they use a softmax function over the set of calculated distances to determine the final set of categories for the test query. Because only embedding-based baselines are included in evaluation, it is unclear how the proposed approach would perform vs.\ traditional IR models.

\textbf{Short text similarity.} To capture the semantic similarity between a pair of short texts,
{\bf \citet{kenter_short_2015}} propose a supervised machine learning algorithm using two different types of meta-features which utilize different publicly available word embeddings. Their first meta-feature resembles BM25, computing semantic text similarity based on word-level semantics (e.g., word embedding of each word). For their second meta-feature, they compute a matrix of cosine similarity in semantic (e.g., word embedding) space between each pair of terms of two sentences, quantizing similarities into three different bins. Along with these two features they introduce two minor text-level features. Each pairwise sentence represented by the features and labeled by the human annotator is used to train a Support Vector Machine (SVM) with Radial Basis Function (RBF) on the MSR Paraphrase Corpus dataset\footnote{\url{https://www.microsoft.com/en-us/download/details.aspx?id=52398}} \cite{dolan2004unsupervised}. Unlike many prior semantic similarity methods, this work requires no external resources (e.g., WordNet), and thus reduces the necessity of human feature engineering. The authors use both \wv{} and GloVe~\cite{pennington2014glove} embeddings 
as well as their own ``auxiliary'' trained embeddings on a dataset from INEX. Results suggest that their method using only publicly available embeddings, without any parameter tuning, outperforms state-of-the-art methods. Use of the auxiliary word embeddings in addition to public embeddings increases performance further. Future work could extend the approach to consider word order.

\section{Background on Neural Networks}
\label{section:nn-background}

Assuming a readership familiar with IR but possibly less versed in neural network (NN) concepts, this section briefly introduces several such concepts which underlie surveyed work in Section~\ref{section:nn}. For further details on these concepts, please see the many informative resources on NN modeling cited in Section~\ref{section:introduction}.

\subsubsection{Neural Network (NN)}
\begin{figure}[ht]
    \centering
    \includegraphics[width=0.3\textwidth]{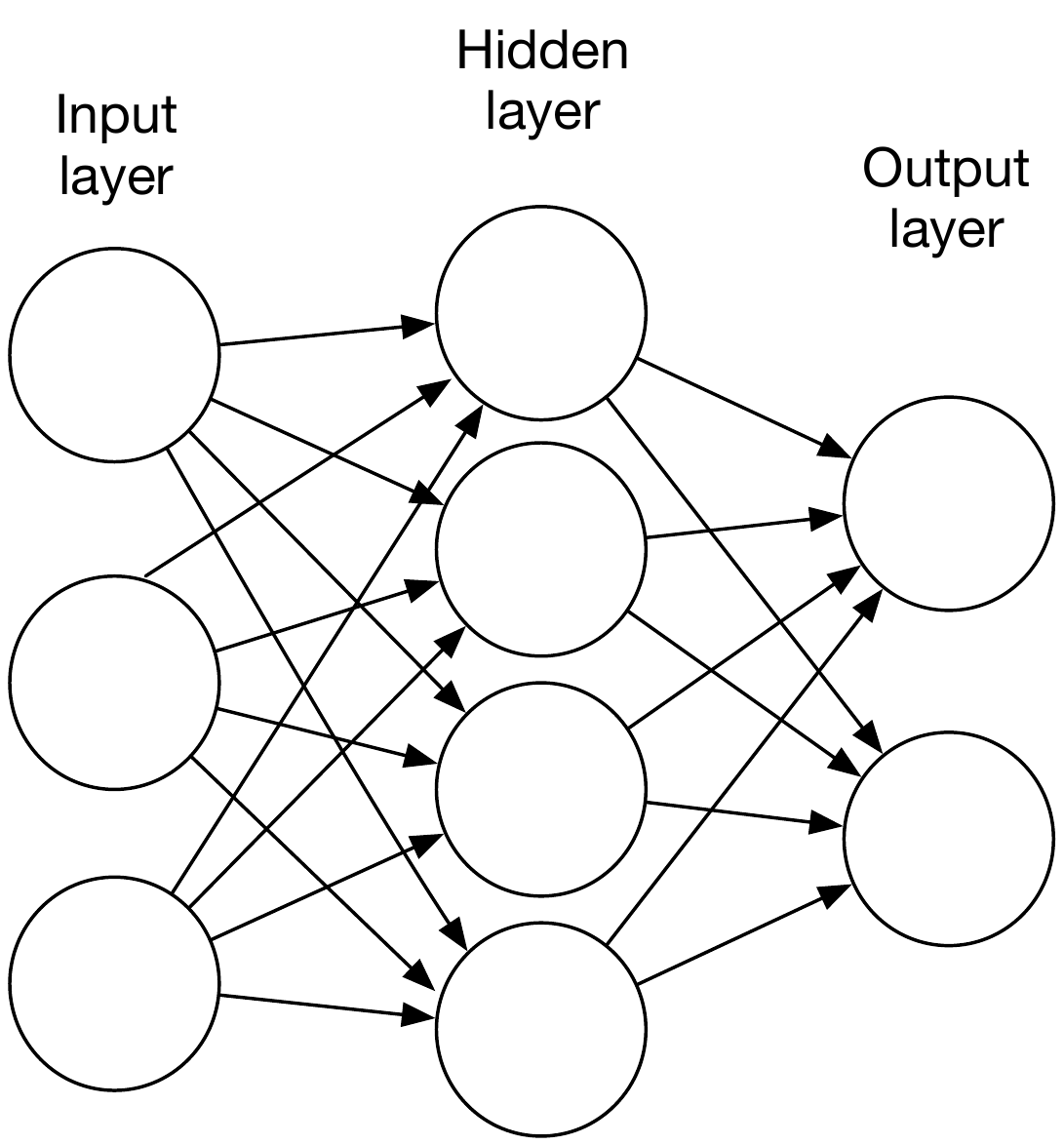} 
    \caption{Feed-forward fully connected neural network}
    \label{figure_NN}
\end{figure}

The neural models we will typically consider in this survey are \emph{feed-forward} networks, which we refer to as neural networks (NNs) for simplicity. A simple example of such an NN is shown in {\bf Figure \ref{figure_NN}}. Input features are extracted, or \emph{learned}, by NNs using multiple, stacked fully connected layers. Each layer applies a linear transformation to the vector output of the last layer (performing an affine transformation). Thus each layer is associated with a matrix of parameters, to be estimated during learning. This is followed by element-wise application of a non-linear activation function. In the case of IR, the output of the entire network is often either a vector representation of the input or some predicted scores. During training, a loss function is constructed by contrasting the prediction with the ground truth available for the training data, where training adjusts network parameters to minimize loss. This is typically performed via the classic back-propagation algorithm \cite{rumelhart1988learning}. For further details, see \citet{goodfellow_deep_2016}. 


\subsubsection{Auto-encoder}

An {\em auto-encoder} neural network is an unsupervised model used to learn a representation for data, typically for the purpose of dimensionality reduction. Different from the typical NN, an auto-encoder is trained to reconstruct the input, and the output has the same dimension as the input. For more details, see \citet{erhan2010does} and \citet{hinton2006reducing}. 

\subsubsection{Restricted Boltzmann Machine (RBM)} A Restricted Boltzmann Machine (RBM) is a stochastic neural network whose binary activations depend on its neighbors and have a probabilistic binary activation function. RBMs are useful for dimensionality reduction, classification, regression, collaborative filtering, feature learning, topic modeling, etc. The RBM was originally proposed by \citet{smolensky1986information} and further popularized by \citet{nair2010rectified}.

\subsubsection{Convolutional Neural Network (CNN)}

\begin{figure}[ht]
	\vspace{-10pt}
    \centering
    \includegraphics[width=0.5\textwidth]{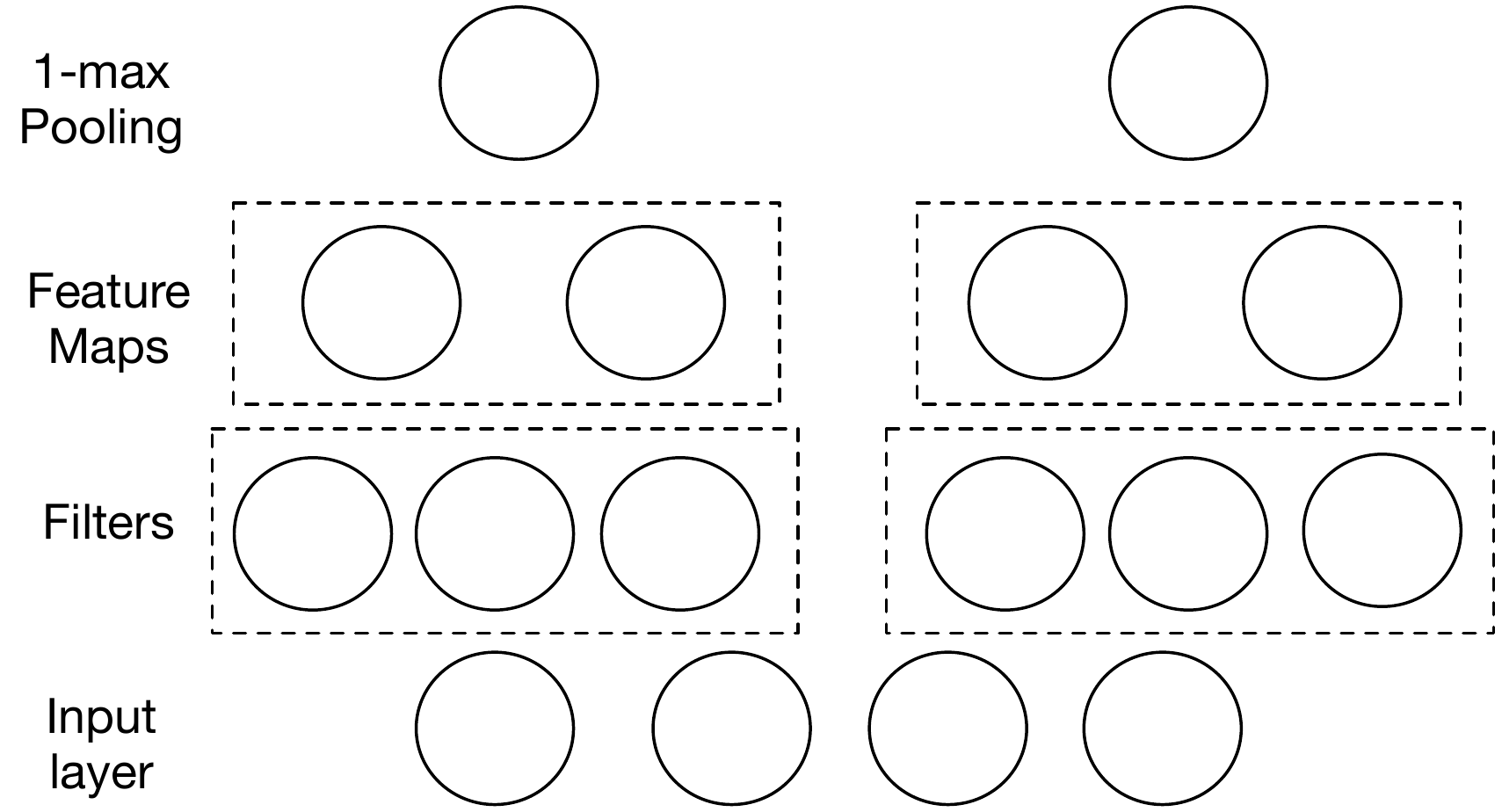} 
    
    \caption{One-dimension Convolution Neural Network with only one convolution layer (two filters and two feature maps), followed by a 1-max-pooling layer.}
    \label{figure_CNN}
\end{figure}
In contrast to the densely-connected networks described above, a Convolutional Neural Network (CNN) \cite{lecun1995convolutional} defines a set of linear filters (kernels) connecting only spatially local regions of the input, greatly reducing computation. These filters extract locally occurring patterns. CNNs are typically built following a ``convolution+pooling'' architecture, where a pooling layer following convolution further extracts the most important features while at the same time reducing dimensionality. We show a basic example of a CNN in {\bf Figure \ref{figure_CNN}}. CNNs were first established by their strong performance on image classification (\citet{krizhevsky2012imagenet}), then later adapted to text-related tasks in NLP and IR (\citet{collobert2011natural, kalchbrenner2014convolutional, kim2014convolutional, zhang2015sensitivity}). As discussed in Section~\ref{section:method-cnn}, \citet{yang_anmm:_2016} and \citet{guo_deep_2016} question whether CNN models developed in computer vision and NLP to exploit spatial and positional information are equally-well suited to the IR domain.

\subsubsection{Recurrent Neural Network (RNN)}

\begin{figure}[ht]
    \centering
    \includegraphics[width=0.3\textwidth]{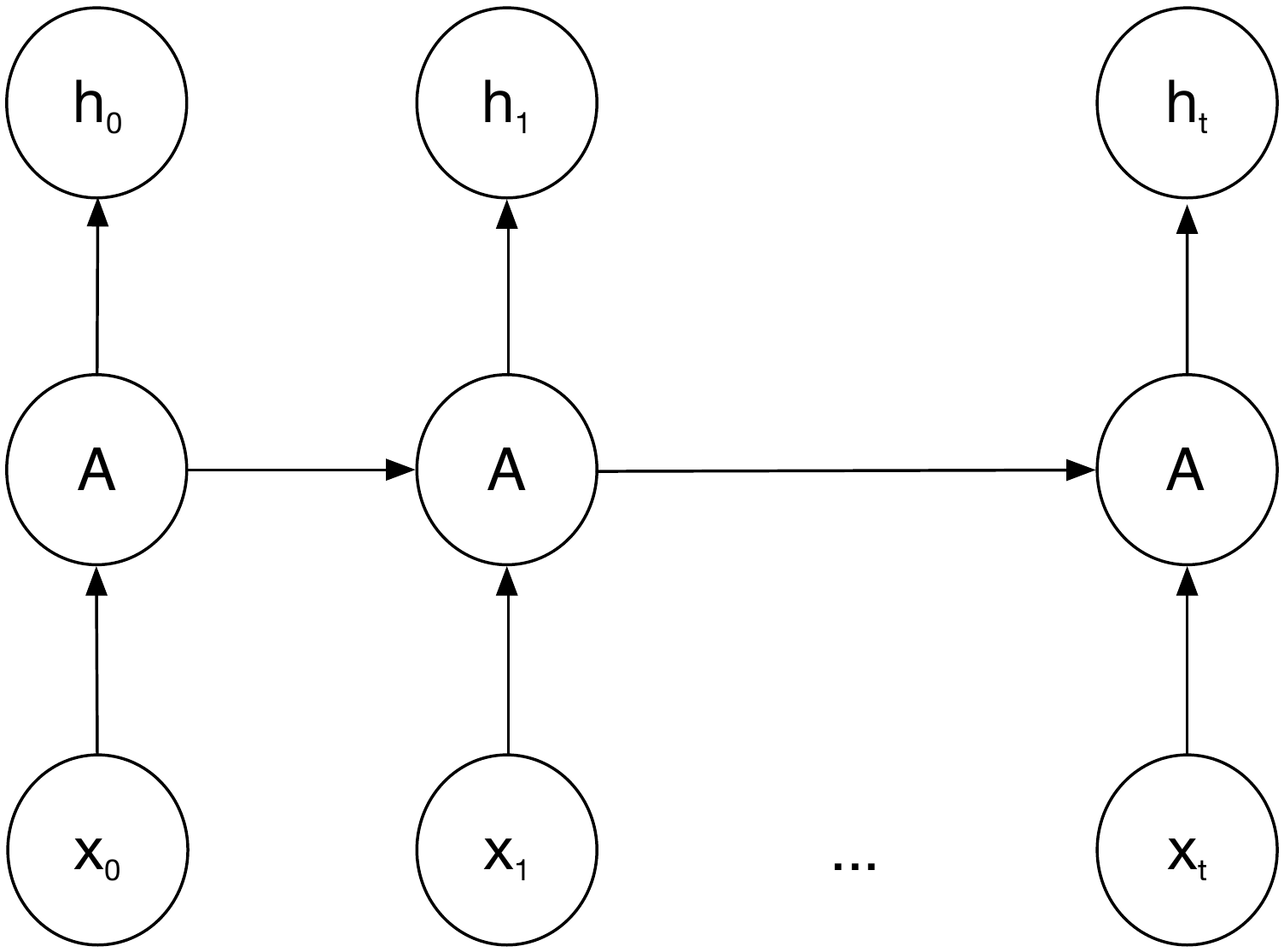} 
    \caption{Recurrent Neural Network. x is the input, A is the computation unit shared across time steps, and h is the hidden state vector}
    \label{figure_RNN}
\end{figure}
A Recurrent Neural Network (RNN) (\citet{elman1990finding}) models sequential input, e.g., sequences of words in a document. We show a basic example of an RNN in {\bf Figure \ref{figure_RNN}}. Individual input units (e.g., words) are typically encoded in vector representations. RNNs usually read inputs sequentially; one can think of the input order as indexing ``time''. Thus the first word corresponds to an observation at time 0, the second at time 1, and so on. The key component of RNNs is a hidden state vector, which encodes salient information extracted from the input read thus far. At each step during traversal (at each time point $t$), this state vector is updated as a function of the current state vector and the input at time $t$. Thus each time point is associated with its own unique state vector, and when the end of a piece of text is reached, the state vector will capture the context induced by the entire sequence. 

A technical problem with fitting the basic RNN architecture just described is the ``vanishing gradient problem'' \cite{pascanu2013difficulty} inherent to parameter estimation via back-propagation ``through time''. The trouble is that gradients must flow from later time steps of the sequence back to earlier bits of the input. This is difficult for long sequences, as the gradient tends to degrade, or ``vanish'' as it is passed backwards through time. Fortunately, there are two commonly used variants of RNN that aim to mitigate this problem (and have proven empirically successful in doing so): LSTM and GRU.

{\bf Long Short Term Memory (LSTM)} (\citet{hochreiter1997long}) was the first approach introduced to address the vanishing gradient problem. In addition to the hidden state vector, LSTM introduces a \emph{memory cell} structure, governed by three gates. An \emph{input gate} is used to control how much the memory cell will be influenced by the new input; a \emph{forget gate} dictates how much previous information in the memory cell will be forgotten; and an \emph{output gate} controls how much the memory cell will influence the current hidden state. 
All three of these gates depend on the previous hidden state and the current input. For a more detailed description of LSTM and more LSTM variants, see \citet{greff2015lstm} and \citet{graves2013generating}.

\citet{bahdanau2014neural}'s {\bf Gated Recurrent Unit (GRU)} is a more recent architecture, similar to the LSTM model but simpler (and thus with fewer parameters). Empirically, GRUs have been found to perform comparably to LSTMs, despite their comparative simplicity \cite{chung2014empirical}. Instead of the memory cell used by LSTMs, an \emph{update gate} is used to govern the extent to which the hidden gate will be updated, and a \emph{reset gate} is used to control the extent to which the previous hidden state will influence the current state.

\subsubsection{Attention}

The notion of \emph{attention} has lately received a fair amount of interest from NN researchers. The idea is to imbue the model with the ability to \emph{learn} which bits of a sequence are most important for a given task (in contrast, e.g., to relying only on the final hidden state vector). 

Attention was first proposed in the context of neural machine translation model by \citet{bahdanau2014neural}. The original RNN model (\citet{sutskever2014sequence}) for machine translation encodes the source sentence into a fixed-length vector (by passing an RNN over the input, as described above). This is then accepted as input by a \emph{decoder} network, which uses the single encoded vector as the only information pertaining to the source sentence. This means all relevant information required for the translation must be stored in a single vector -- a difficult aim. 

The attention mechanism was proposed to alleviate this requirement. At each time step (as the decoder generates each word), the model identifies a set of positions in the source/input sentence that is most relevant to its current position in the output (a function of its index and what it has generated thus far). These positions will be associated with corresponding state vectors. The current contextualizing vector (to be used to generate output) can then be taken as a sum of these, weighted by their estimated relevance. This attention mechanism has also been used in image caption generation (\citet{xu2015show}). A similar line of work includes \emph{Neural Turing Machines} by \citet{graves2014neural} and \emph{Memory Networks} by \citet{weston2014memory}.



\section{Neural Network Approaches to IR}
\label{section:nn}

Availability of code and existing embeddings from \wv{}~\cite{mikolov2013distributed} and GloVe~\cite{pennington2014glove} (Tables \ref{table:code} and \ref{table:data}, respectively) have provided a key avenue for work on Neural IR, especially for extending traditional IR models to integrate word embeddings. Section \ref{section:word-embedding} surveyed studies in this space. In contrast, 
we now survey conceptually different approaches which directly incorporate word embeddings within NN models, reflecting a more significant shift toward pursuing {\em end-to-end} NN architectures in IR. We organize surveyed studies by IR task (see {\bf Table \ref{Tasks_NN}}). For additional background on NN concepts underlying the approaches surveyed in this section, see Section \ref{section:nn-background}.

\begin{figure}[ht]
	\vspace{-10pt}
	\centering
    \includegraphics[width=\textwidth]{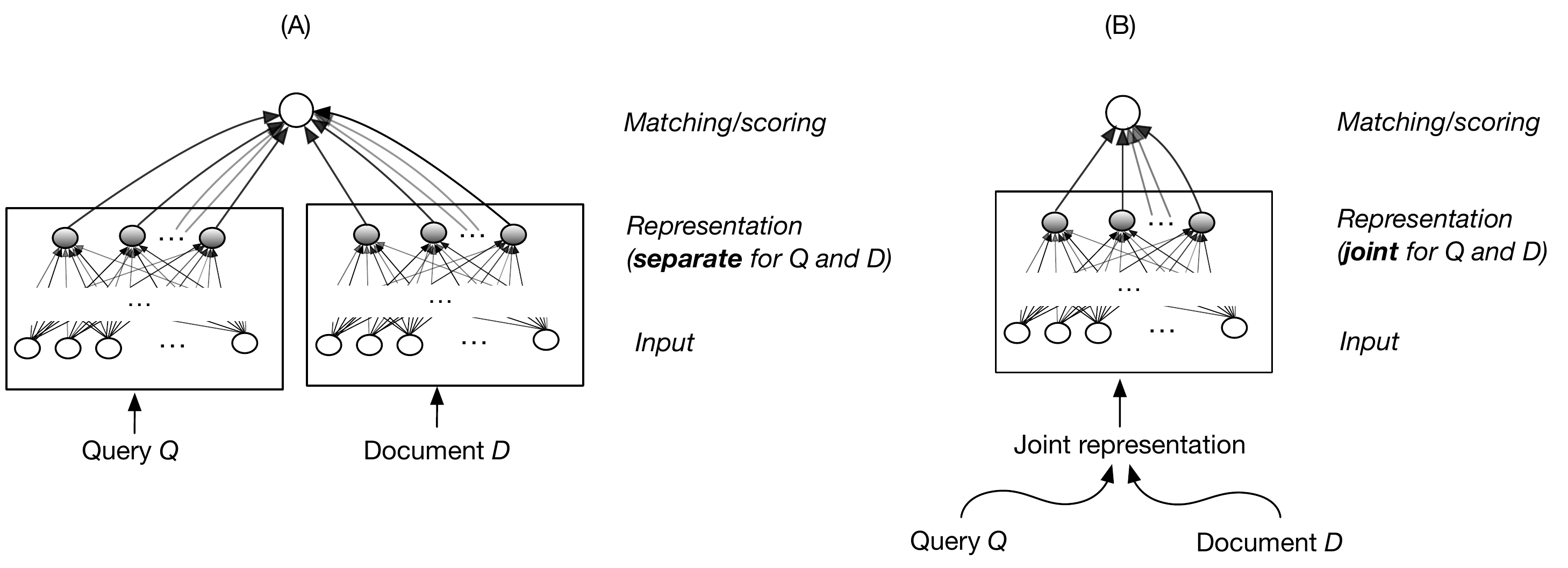}
    \caption{Two basic neural architectures for scoring the relevance of queries to documents. On the left (A), the query and document are independently run through mirror neural models (typically this would be a Siamese architecture \cite{bromley_siamese_1993}, in which weights are shared between the networks); relevance scoring is then performed by a model operating over the two induced representations. On the right (B), one first constructs a joint representation of the query and document pair and then runs this joint input through a network. The architectures differ in their inputs and the semantics of the hidden/representational layers. This figure is inspired by Figure 1 in Guo et al. \cite{guo_deep_2016}.}
    \label{fig:two-approaches}
\end{figure}



{
\renewcommand{\arraystretch}{1.3}
\begin{table}[ht]
\centering
  \begin{tabular}
  {| p{3.0cm}| p{12.0cm} |}
    \hline    
    {\bf Task} & {\bf Studies} \\ \hline
	\Adhoc{} Retrieval &  BP-ANN~(\citet{yang_selective_2016}), CDNN~(\citet{severyn_learning_2015}), C-DSSM~(\citet{shen_learning_2014}), CLSM~(\cite{shen_latent_2014}), DSSM~(\citet{huang_learning_2013}), DRMM~(\citet{guo_deep_2016}), GDSSM~(\citet{ye_learning_2015}), \citet{gupta_query_2014}, \citet{li_deep_2014}, \citet{nguyen_toward_2016}, QEM (\citet{sordoni_learning_2014}) \\
    \hline
    Conversational Agents & DL2R~(\citet{yan_learning_2016}) \\
    \hline
    Proactive Search & \citet{luukkonen_lstm-based_2016} \\
    \hline
    Query Auto-completion & \citet{mitra_exploring_2015,mitra_query_2015} \\
    \hline
    Query Suggestion & \citet{sordoni_hierarchical_2015} \\
    \hline
    Question Answering & BLSTM~(\citet{wang2015long}), CDNN~(\citet{severyn_learning_2015}), DFFN~(\citet{suggu_deep_2016}), DL2R~(\citet{yan_learning_2016}), \citet{yu2014deep} \\
    \hline
    Recommendation & \citet{gao_modeling_2014}, \citet{song_multi-rate_2016} \\
    \hline
    Related Document Search & \citet{salakhutdinov_semantic_2009} \\
    \hline
    Result Diversification & \citet{xu_modeling_2016} \\
    \hline
    Sponsored Search & \citet{zhang_deep_2016} \\
    \hline
    Summarizing Retrieved Documents & \citet{lioma_deep_2016} \\
    \hline
    Temporal IR & \citet{kanhabua_learning_2016} \\
    \hline
  \end{tabular}
  \caption{IR tasks solved by Neural Network methods.}
  \label{Tasks_NN}
\end{table}
}

\subsection{\Adhoc{} Retrieval} 

As discussed by \citet{guo_deep_2016}, neural models for this tend to follow one of two general architectures, which we depict in {\bf Figure \ref{fig:two-approaches}}. In the first (Figure \ref{fig:two-approaches}A), queries and documents are separately passed through neural networks (with shared parameters) and scores are generated based on the combination (e.g., concatenation) of the respective outputs. By contrast, Figure \ref{fig:two-approaches}B depicts an approach in which a joint representation for queries and documents is constructed and this is run through a neural network. \citet{guo_deep_2016}'s work is further discussed below.

{\bf \citet{huang_learning_2013}} propose a Deep Structured Semantic Model (DSSM) for \adhoc{} web search. DSSM was one of the pioneer models incorporating click-through data in deep NNs and has been built on by a variety of others \cite{shen_learning_2014,shen_latent_2014,ye_learning_2015,mitra_exploring_2015,mitra_query_2015}. Notably, documents are indexed only by title text rather than the entire body text. The query and the document are first modeled as two high dimensional term vectors (i.e., a bag-of-words representation). DSSM learns a representation of documents and queries via a feed-forward NN to obtain a low-dimensional vector projected within a latent semantic space. The document ranking is then trained within the DSSM architecture by maximizing the conditional likelihood of query given document. More specifically, the authors estimate this conditional likelihood by a softmax function applied on the cosine similarity between the corresponding semantic vector of documents and queries. Given large vocabularies and the need to support large-scale training, the authors propose a word hashing method which transforms the high-dimensional term vector of the query/document to a low-dimensional letter-trigram vector. This lower dimensional vector is then considered as input to the feed-forward NN. Trigram hashing also helps to address out of vocabulary (OOV) query terms not seen in training data. The authors do not discuss how to mitigate hashing collisions; while they show that such collisions are relatively rare (e.g., $0.0044\%$ for 500K vocabulary size), this stems in part from indexing document titles only and not body text. Later work by \citet{guo_deep_2016} performs two further experiments with DSSM: indexing only document titles vs.\ indexing entire documents (i.e., full-text search). \citet{guo_deep_2016}'s results indicate that full-text search with DSSM does not perform as well as traditional IR models. 

{\bf \citet{shen_learning_2014}} extend DSSM \cite{huang_learning_2013} by introducing a CNN with max-pooling in the DSSM architecture (C-DSSM). It first uses word hashing to transform each word into a vector. A convolutional layer then projects each word vector within a context window to a local contextual feature vector. It also incorporates a max-pooling layer to extract the most salient local features to form a fixed-length global feature vector for queries and web documents. 
The main motivation for the max-pooling layer is that because the overall semantic meaning of a sentence is often determined by a few key words, simply mixing all words together (e.g., by summing over all local feature vectors) may introduce unnecessary divergence and hurt the overall semantic representation effectiveness. This is a key difference between DSSM and C-DSSM. 

Both DSSM \cite{huang_learning_2013} and C-DSSM \cite{shen_learning_2014}
fail to capture the contextual information of the queries and the documents. To address this, {\bf \citet{shen_latent_2014}} propose a Convolutional Latent Semantic Model (CLSM) built atop DSSM.
CLSM captures the contextual information by a series of projections from one layer to another in a CNN architecture \cite{lecun1995convolutional}. The first layer consists of a word n-gram followed by a letter trigram layer where each word n-gram is composed of its trigram representation, a form of the word hashing technique developed in DSSM. Then, a convolution layer transforms these trigrams into contextual feature vectors using the convolution matrix $W_c$, which is shared among all the word n-grams. Max-pooling is then performed against each dimension on a set of contextual feature vectors. This generates the global sentence level feature vector $v$. Finally, using the non-linear transformation $tanh$ and the semantic projection matrix $W_s$, they compute the final latent semantic vector for the query/document. 
The parameters {$W_c$ and $W_s$} are optimized to maximize the same loss function used by \citet{huang_learning_2013} for DSSM. Even though CLSM introduces the word n-gram to capture the contextual information, it suffers the same problems as DSSM including scalability. For example, CLSM performs worse when trained on a whole document than when trained on only the document title, as shown in experiments reported by \citet{guo_deep_2016}. 

{\bf \citet{nguyen_toward_2016}} propose two high level views of how to incorporate a knowledge base (KB) graph into a ranking model like DSSM \cite{huang_learning_2013}. To the best of our knowledge, this is one of the first IR studies which tries to incorporate KB into deep neural structure. The authors' first model exploits knowledge bases to enhance the representation of a document-query pair and its similarity score by exploiting concept embeddings learned from the KB distributed representation \cite{xu2014rc,faruqui2014retrofitting} as input to deep NNs like DSSM. The authors argue that this hybrid representation of the distributional semantic (namely, word embeddings) and the symbolic semantics (namely, concept embeddings taking into account the graph structure) would enhance document-query matching. Their second model uses the knowledge resource is as an intermediate component that helps to translate the deep representation of the query towards the deep representation of the document with respect to the \adhoc{} IR task. Strong empirical evidence is still needed to demonstrate that adding KB indeed benefits the deep neural architecture for capturing semantic similarity. 

Although DSSM \cite{huang_learning_2013}, C-DSSM \cite{shen_learning_2014} and CLSM \cite{shen_latent_2014} achieve superior performance over traditional latent semantic models (e.g., PLSA, LDA), these models make some strong assumptions about clicked $\langle query,document\rangle$ pairs. According to {\bf \citet{ye_learning_2015}} these three assumptions are: i) each clicked $\langle query,document\rangle$ pair is equally weighted; 
ii) each clicked $\langle query,document\rangle$ pair is a 100\% positive example;
and iii) each click is solely due to the semantic similarity. The authors relax these assumption and propose two generalized extensions to DSSM: 
GDSSM1 and GDSSM2. While DSSM models the probability of a document being clicked given a query and the semantic similarity between document and query, GDSSM1 uses more information in its loss function, incorporating the number of users seeing and proportion clicking for each query-document pair. GDSSM2 conditions on lexical similarity in addition to semantic similarity. Potential future work includes developing a similar loss function as used in CLSM \cite{shen_learning_2014}.  

{\bf \citet{li_deep_2014}} propose a document re-ranking model based on fine-grained features derived from user contextual data, such as the history of user queries and clicked documents. The major contribution of their work is to model user dissatisfaction from the click-through data and develop semantic similarity between previous and current queries based upon this dissatisfaction. The authors model contextual information as a combination of 
``global'' (i.e., query-level) and ``local'' (single URL-level) components. Their evaluation involves a re-ranking task in which the rank of an original search engine is used as an input feature alongside the contextual features derived by their neural models. They evaluate XCode (internally developed by the authors), DSSM \cite{huang_learning_2013}, and C-DSSM \cite{shen_learning_2014} as models for deriving these contextual features and find that DSSM offers the highest performance, followed by C-DSSM. Though they expected C-DSSM to offer the highest performance, they note that C-DSSM could only be trained on a small dataset with bounded size, in contrast to DSSM, which could be trained on a larger dataset. Additionally, they note that the performance difference could be due to imperfect tuning of model parameters, such as sliding window size for C-DSSM. Nevertheless, contextual features derived using both DSSM and C-DSSM offer performance benefits for re-ranking. 
Potential future work includes incorporating user profiles in re-ranking. 

{\bf \citet{guo_deep_2016}}'s Deep Relevance Matching Model (DRMM) is one of the first NN IR models to show improvement over traditional IR models (though comparison is against bag-of-words approaches rather than any term proximity baselines, such as \citet{metzler2005markov}'s Markov Random Field model). The authors hypothesize that deep learning methods developed and/or commonly applied in NLP for semantic matching may not be well-matched with \adhoc{} search, which is most concerned with relevance matching. They articulate three key differences they perceive between semantic and relevance matching:
\begin{enumerate}
\item Semantic matching looks for semantic similarity between terms; relevance matching puts more emphasis on exact matching.
\item Semantic matching is often concerned with how composition and grammar help to determine meaning; varying importance among query terms is more crucial than grammar in relevance matching.
\item Semantic matching compares two whole texts in their entirety; relevance matching might only compare parts of a document to a query.
\end{enumerate}
The raw inputs to the DRMM are term embeddings. The first layer consists of \textit{matching histograms} of each query term's cosine similarity scores with each of the document terms. On top of this histogram layer is an NN outputting a single node for each query term. These outputs are multiplied by query term importance weights calculated by a \textit{term gating network} and then summed to produce the final predicted matching score for the query/document pair. The whole network is trained using a margin ranking loss function.   
\Adhoc{} search experiments are reported on TREC Robust04 and \clueweb{}. Baselines include: i) traditional retrieval models such as BM25 and query likelihood; ii) representation-focused NN models, including DSSM \cite{huang_learning_2013} and C-DSSM \cite{shen_learning_2014} (indexing titles vs.\ full documents), ARC-I \cite{hu_convolutional_2014}; and iii) interaction-focused NN models, such as ARC-II \cite{hu_convolutional_2014} and MatchPyramid \cite{pang_text_2016}. OOV terms are represented by random vectors (as in \cite{kenter_short_2015}), effectively allowing only exact matching. Results show that as CBOW dimension increases (50, 100, 300 and 500), DRRM first increases then slightly drops. DRRM using IDF and Log-Count Histogram (LCH) also significantly outperforms all baselines. In addition, all representation-focused and interaction-focused baselines are not competitive with traditional retrieval models, supporting the authors' hypothesis that deep matching models focused on semantic matching may not be well-suited to \adhoc{} search. For future work, they propose using click-through log data to augment training.

{\bf \citet{yang_selective_2016}} propose a document ranking model which decides if using term proximity ranking for a query would be beneficial based on its features. Term proximity (TP) requires especially large indexes and is computationally expensive. The proposed model seeks to return improved rankings while also reducing the overhead incurred by using TP only when it can be helpful. \citet{yang_selective_2016} use the popular TP ranking functions BM25TP \cite{rasolofo2003term}, Markov Random Field (MRF) \cite{metzler2005markov}, and ``EXP'' \cite{tao2007exploration}. 
If a query achieves better MAP results using BM25TP or EXP over BM25 or using MRF over TF \cite{metzler2005markov}, its features are labeled 1, otherwise 0. The results are used to train a Back-Propagation Artificial Neural Network (BP-ANN) \cite{rumelhart1988learning}, which is used to build the proposed selective model. 
A combination of statistical methods (ranksum, z-score, $\chi$ -square), a decision tree, and a feature weight algorithm (relief) are employed to find important features for BM25TP, MRF, and EXP to construct the BP-ANN. 
The experimental evidence shows that consistently using TP achieves significantly better rankings. However, the results suggest that selectively using TP returns slightly better rankings and has better throughput than when always using TP. Future work could include introducing more important features capturing term proximity, as well as training a collection of models, each designed for a particular query type, for query-dependent feature generation.

\subsubsection{Query Expansion}

{\bf \citet{sordoni_learning_2014}} propose a supervised Quantum Entropy Minimization (QEM) approach for finding semantic representation of {\em concepts}, such as words or phrases, for query expansion. The authors suggest that text sequences should not lie in the same semantic space as single terms because their information content is higher. To this end, concepts are embedded in rank-one matrices while queries and documents are embedded as a mixture of rank-one matrices. This allows documents and queries to lie in a larger space and carry more semantic information than concepts, thereby achieving greater semantic resolution. As such, QEM strives for an increased representational power vs.\ existing approaches relying only on vector representations (for example, combining individual word embeddings by simple arithmetic operations like averaging to learn semantic representations for larger textual units; see broader discussion in Section~\ref{section:method-representation}). For learning parameters of {\em density matrices}, QEM's gradient updates are a refinement of updates in both ~\citet{bai2009supervised}'s Supervised Semantic Indexing (SSI) and Rocchio~\cite{rocchio1971relevance}; a query concept embedding moves toward its nearest relevant documents’ concepts and away from its nearest non-relevant documents’ concepts. This has the effect of selecting which document concepts the query concept should be aligned with and also leads to a refinement of Rocchio: the update direction for query expansion is obtained by weighting relevant and non-relevant documents according to their similarity to the query. Due to lack of public query logs, QEM is trained on Wikipedia {\em anchor logs}~\cite{dang2010query}. 
Experiments conducted on \clueweb{} with TREC 2010-2012 Web Track topics show QEM outperforms \citet{gao2010clickthrough}'s concept translation model (CTM) (statistically significant differences), and SSI (occasionally statistically significant differences). QEM notably achieves improved precision at top-ranks. 
Also notable is QEM's ability to find useful expansion terms for longer queries due to higher semantic resolution. Additional preliminary experiments with \citet{weston2010large}'s \textit{Weighted Approximate-Rank Pairwise} loss yielded further improvements for QEM over baselines. Proposed future work includes exploring ways to automatically set appropriate embedding dimensions.  In addition, the difficult maximization when computing the embeddings for queries and documents relies on a linear Taylor expansion for approximation, which might be further improved. Better approximations may exist than the linear approximation used here, particularly for the scoring function.


{\bf \citet{gupta_query_2014}} explore query expansion in mixed-script IR (MSIR), a task in which documents and queries in non-Roman written languages can contain both native and transliterated scripts together. Stemming from the observation that transliterations generally use Roman letters in such a way as to preserve the original-language pronunciation, the authors develop a method to convert both scripts into a bilingual embedding space. Therefore, they convert terms into feature vectors of the count of each Roman letter. An {\em auto-encoder} is then trained for query expansion by feeding in the concatenated native term feature vector and transliterated term feature vector. Experiments are conducted on shared task data from FIRE (Section~\ref{section:we-task-adhoc}).
Results suggest that their method significantly outperforms other state-of-the-art methods. 
For future work, while the query projection into abstract space is performed here using matrix multiplication online, this could be further optimized. An ablation study of features would help elucidate the relative importance of different features (abstract feature space) and to determine which contribute most to model performance. Source code for the model is shared online (see Table~\ref{table:data}). 
%


\subsection{Query Auto-completion}
{\bf \citet{mitra_exploring_2015}} learns embedding vectors for query reformulation based on query logs, representing query reformulation as a vector using \citet{shen_learning_2014}'s  Convolution Latent Semantic Model (CLSM). Two sets of features are developed. The first feature set captures topical similarity via cosine similarity between the candidate embedding vector and embedding vectors of some past number of queries in the same session. The second set of \emph{reformulation features} captures the difference between the candidate embedding vector and that of the immediately preceding query. Other features used include non-contextual features, such as most popular completion, as well as contextual features, such as n-gram similarity features and pairwise frequency features. LambdaMART \cite{wu2010adapting} is trained on click-through data and features. Results suggest that embedding features give considerable improvement over Most Probable Completion (MPC) \cite{bar2011context}, in addition to other models lacking the embedding-based features. 

Whereas traditional query auto-completion (QAC) methods perform well for head queries having abundant training data, prior methods often fail to recommend completions for tail queries having less usual prefixes, including both probabilistic algorithms \cite{bar2011context,cai2014time, shokouhi2012time} and learning-based QAC approaches \cite{cai2016learning,jiang2014learning}. To address this, {\bf \citet{mitra_query_2015}} develop a query auto-completion procedure for such rare prefixes which enables QAC even for query prefixes never seen in training. The authors mine the most popular query suffixes (e.g., n-grams which appear at the end of a query) and append them to the user query prefixes, thus generating possible synthetic candidate solutions. To recommend possible query expansion, they use LambdaMART \cite{wu2010adapting} with two sets of ranking features. The first feature set is the frequency of query n-grams in the search log. The second feature set is generated by training CLSM on the prefix-suffix dataset, sampling queries in the search logs and segmenting each query at every possible word boundary. 
Results on the AOL search log show significant improvement over MPC \cite{bar2011context}. The authors also find n-gram features to be more important than CLSM features in contributing to model performance. 


\subsection{Question Answering: Answer Sentence Selection}

Methods proposed in this section all evaluate on \citet{wang2007jeopardy}'s dataset derived from TREC QA 8-13. 

{\bf \citet{wang2015long}} 
use a stacked bi-directional LSTM (BLSTM) to read a question and answer sequentially and then combine the hidden memory vectors from LSTMs of both question and answer. They use mean, sum and max-pooling as features. This model needs to incorporate key-word matching as a feature to outperform previous approaches that do not utilize deep learning. They use BM25 as the key-word matching feature and use Gradient boosted regression tree (GBDT) \citet{friedman2001greedy} to combine it with the LSTM model.  
 
{\bf \citet{severyn_learning_2015}} propose a Convolutional Deep Neural Network (CDNN) to rank pairs of short texts. Their deep learning architecture has 2 stages. The first stage is a sentence embedding model using a CNN to embed question and answer sentences into intermediate representative vectors, which are used to compute their similarity. The second stage is a NN ranking model whose features include intermediate representative sentence vectors, similarity score, and some additional features such as word overlap between sentences. Results show improvement of about 3.5\% in MAP vs.\ results reported by \citet{yu2014deep}, in which a CNN is used followed by logistic regression (LR) to rank QA pairs. The authors attribute this improvement to the larger width (of 5) of the convolutional filter in their CNN for capturing long term dependencies, vs.\ the unigram and bigram models used by \citet{yu2014deep}. Beyond the similarity score, their second stage NN also takes intermediate question and answer representations as features to constitute a much richer representation. \citet{yu2014deep} do not include such intermediate question and answer representations as features to their LR approach.

{\bf \citet{yang_anmm:_2016}}'s approach starts by considering the interaction between question and answer at the word embedding level. They first build a question-answer interaction matrix using pre-trained embeddings. They then use a novel value-shared weight CNN layer (instead of a position-shared CNN) in order to induce a hidden layer. The motivation for this is that different matching ranges between a question term and answer will influence the later ranking score differently. After this, they incorporate an {\em attention network} for each question term to explicitly encode the importance of each question term and produce the final ranking score. They rank the answer sentences based on the predicted score and calculate MAP and MRR. Whereas \citet{severyn_learning_2015} and \citet{wang2015long} need to incorporate additional features in order to achieve comparative performance, \citet{yang_anmm:_2016} do not require any feature engineering. 

\subsection{Community Question Answering (CQA)}
{\bf \citet{suggu_deep_2016}} propose a Deep Feature Fusion Network (DFFN) to exploit both hand-crafted features (HCF) and deep learning based systems for Answer Question Prediction. Specifically, query/answer sentence representations are embedded using a CNN. A single feature vector of 601 dimensions serves as input to a second stage fully-connected NN. Features include sentence representations, HCF (e.g., TagMe, Google Cross-Lingual Dictionary (GCD), and Named Entities (NEs)), similarity measures between questions and answers, and metadata such as an answer author's reputation score. The output of the second stage NN is a score predicting the answer quality. They compare their approach to the top two best performing HCF based systems from SemEval 2015 and a pure deep learning system. For SemEval 2016, DFFN was compared with their corresponding top two best performing system. Results show that DFFN performs better than the top systems across all metrics (MAP, F1 and accuracy) in both SemEval 2015 and 2016 datasets. The authors attribute this to fusing the features learned from deep learning and HCF, since some important features are hard to learn automatically. As an example, question and answer often consists of several NEs along with variants, which are hard to capture using deep learning. However, NEs can be extracted from QA and their similarity used as a feature. Their use of HCF was also motivated by the many available similarity resources, such as Wikipedia, GCD, and click-through data, which could be leveraged to capture richer syntactic and semantic similarities between QA pairs.

\subsection{Recommendation}
Due to difficulty modeling temporal behavior in recommendation systems, many existing techniques treat users' interests as static long-term features. {\bf \citet{song_multi-rate_2016}} are mainly concerned with how to incorporate short-term users' interest to improve the recommendation quality of news, where freshness is particularly important. Based on DSSM~\cite{huang_learning_2013}, they propose a temporal deep semantic structured model (TDSSM). It has two views represented by two neural networks. The item view contains the ``implicit feedback of items'' and the user view contains ``user's query history.'' An RNN is used to model short-term temporal user interests. They also expand their model to a multi-rate-TDSSM (MR-TDSSM) model, which can take different granularities of time as input: slow-rate RNNs take daily (user, news) clicks, whereas fast-rate RNNs take weekly clicks. Data used consists of (user, news) clicks from a commercial News recommendation system of user news click history over six months. They compare the performance of TDSSM with several traditional recommendation algorithms and find that TDSSM and MR-TDSSM outperform previous methods not using deep learning. 
Regarding future work, the authors propose to apply the proposed model to different recommendation tasks and explore an attention-based memory network model. 

{\bf \citet{gao_modeling_2014}} propose using DSSM for both automatic highlighting of relevant keywords in documents and recommendation of alternative relevant documents based upon these keywords. They evaluate their framework based on what they call \textit{interestingness} tasks, derived from Wikipedia anchor text and web traffic logs. They find that feeding DSSM derived features into a supervised classifier for recommendation offers state-of-the-art performance and is more effective than simply computing distance in the DSSM latent space. 
Future work could incorporate complete user sessions, since prior browsing and interaction history recorded in the session provide useful additional signals 
for predicting interestingness. This signal might be modeled most easily by using an RNN. 

\subsection{Other Tasks}
\label{section:nn-other-tasks}

\textbf{Conversational Agents.}
An automatic conversation response system called Deep Learning-to-Respond (DL2R) is proposed by {\bf \citet{yan_learning_2016}}. They train and test on 10 million $\langle posting, reply\rangle$ pairs of human conversation web data from various sources, including microblog websites, forums, Community Question Answering (CQA) bases, etc. For a given query they reformulate it using other contextual information and retrieve the most likely candidate reply. They model the total score as a function of three scores: query-reply, query-posting, and query-context, each fed into a neural network consisting of bi-directional LSTM RNN layers, convolution and pooling layers, and several feed-forward layers. The strength of DL2R comes from the incorporation of reply, posting, and context with the query. 
However, since DL2R is a supervised learning algorithm and training data is very scarce in practice, a possible future direction would be to develop an unsupervised method for calculating similarity of sentence-pairs.

\textbf{Proactive Search}. In the only work we know of investigating Neural IR for proactive retrieval, {\bf \citet{luukkonen_lstm-based_2016}} propose LSTM-based text prediction for query expansion. Intended to better support proactive intelligent agents such as Apple's {\em Siri}, {\em Ok Google}, etc., the LSTM is used to generate sequences of words based on all previous words written by users. A beam search is used to prune out low probability sequences. Finally, words remaining in the pruned tree are used for query expansion. Because LSTMs have the ability to represent a given text as well as to predict text, the proposed model outperforms traditional models \cite{glowacka2013directing}. The authors provide several possible future directions, such as using the model to automatically suggest different continuations for user text as it is written, as done in the Reactive Keyboard \cite{darragh1990reactive} and akin to query auto-completion in search engines. User studies are also needed to test the system's effectiveness in the context of users' real-world tasks.

\textbf{Query Suggestion}. 
We know of no work using RNNs for query suggestion prior to {\bf \citet{sordoni_hierarchical_2015}}'s training a hierarchical GRU model to generate context-aware suggestions. They first use a GRU layer to encode the queries in each session into vector representations, then build another GRU layer on sequences of query vectors in a session and encode the session into a vector representation. The model learns its parameters by maximizing the log-likelihood of observed query sessions. To generate query suggestions, the model uses a GRU decoder on each word conditioned on both the previous words generated and the previous queries in the same session. The model estimates the likelihood of a query suggestion given the previous query. 
A learning-to-rank system is trained to rank query suggestions, incorporating the likelihood score of each suggestion as a feature. Results on the AOL query log show that the proposed approach outperforms several baselines which use only hand-crafted features. 
The model is also seen to be robust when the previous query is noisy. Moreover, 
because the model can generate synthetic queries, it can effectively handle long tail queries. However, only previous queries from the same session are used to provide the contextual query suggestion; the authors do not utilize click-through data from previous sessions. Because click-through data provides important feedback for synthetic query suggestion, incorporating such click-through data from previous sessions represents a possible direction for future work. 

\textbf{Related Document Search}. Semantic hashing is proposed by {\bf \citet{salakhutdinov_semantic_2009}} to map semantically similar documents nearby to one another in hashing space, facilitating easy search for similar documents. 
Multiple layers of Restricted Boltzmann Machines (RBMs) are used to learn the semantic structure of documents. The final layer is used as a hash code that compactly represents the document. 
The lowest layer is simply word-count data and is modeled by the Poisson distribution. The hidden layers are binary vectors of lower dimensions. 
The deep generative model is learned by first pre-training the RBMs one layer at a time (from bottom to top). The network is then ``unrolled'', i.e., the layers are turned upside down and stacked on top of the current network. 
The final result is an {\em auto-encoder} that learns a low-dimensional hash code from the word-count vector and uses that hash code to reconstruct the original word-count vector. The auto-encoder is then {\em fine-tuned} by back-propagation. 
Results show that semantic hashing is much faster than locality sensitive hashing \cite{andoni2006near, datar2004locality} and can find semantically similar documents in time independent of document collection size. However, it imposes 
difficult optimization and a slow training mechanism, reducing applicability to large-scale tasks \cite{liu2012supervised} and suggesting possible future work. In addition, because relevance judgments were not available, document corpora with category labels were used instead, assuming relevance if a query and document had matching category labels.

\textbf{Result Diversification.}
Prior state-of-the-art methods for diversifying search results include the Relational Learning-to-Rank framework (R-LTR) \cite{zhu2014learning} and the Perceptron Algorithm using Measures as Margins (PAMM) \cite{xia2015learning}. These prior methods either use a heuristic ranking model based on a predefined document similarity function, or they automatically learn a ranking model from predefined novelty features often based on cosine similarity. In contrast, {\bf \citet{xu_modeling_2016}} takes automation a step further, using Neural Tensor Networks (NTNs) to learn the novelty features themselves. The NTN architecture was first proposed to model the relationship between entities in a knowledge graph via a bilinear tensor product \cite{socher2013reasoning}. The model here takes a document and a set of other documents as input. The architecture uses a tensor layer, a max-pooling layer, and a linear layer to output a document novelty score. The NTN augmentations of R-LTR and PAMM perform at least as well as those baselines, showing how the NTN can remove the need for manual design of functions and features. It is not clear whether using a full tensor network works much better than just using a single slice of the tensor. 

\textbf{Sponsored Search.}
Predicting click-through rate (CTR) and conversion rate from categorical inputs such as region, ad slot size, user agent, etc., is important in sponsored search. {\bf \citet{zhang_deep_2016}} propose the first neural approach we know of to predict CTR for advertising. An interesting aspect of the work is learning vector embeddings of categorical variables in a similar way to natural language terms (see \citet{zhou_learning_2015} for another example of integrating categorical data with word embeddings). The authors develop two deep learning approaches to this problem, a Factorization Machine supported Neural Network (FNN) and Sampling-based Neural Network (SNN). The Factorization Machine is a non-linear model that can efficiently estimate feature interactions of any order even in problems with high sparsity by approximating higher order interaction parameters with a low-rank factorized parameterization. The use of FM-based bottom layer in the deep network, therefore, naturally solves the problem of high computational complexity of training neural networks with high-dimensional binary inputs. The SNN is augmented either by a sampling-based Restricted Boltzmann Machine (SNN-RBM) or a sampling-based Denoising Auto-Encoder (SNN-DAE). The main challenge is that given many possible values of several categorical fields, converting them into dummy variables results in a very high-dimensional and sparse input space. For example, thirteen categorical fields can become over 900,000 binary inputs in this problem. The FNN and SNN reduce the complexity of using a neural network on such a large input by limiting the connectivity in the first layer and by pre-training by selective sampling, respectively. After pre-training, the weights are {\em fine-tuned} in a supervised manner using back-propagation. Evaluation focuses on the tuning of SNN-RBM and SNN-DAE models and their comparison against logistic regression, FM and FNN, on the {\em iPinYou} dataset\footnote{\url{http://data.computational-advertising.org}} \cite{liao2014ipinyou} with real user click data, divided into five test sets. Results on each test set show one of the proposed methods performs best, though the baselines are often close behind and twice take second place. The authors also find a diamond-shape architecture is better than increasing, decreasing, or constant hidden-layer sizes and that dropout works better than L2 regularization. For future work, the model performance might be improved by a momentum method for handling the curvature problems in the DNN training objectives without using complex second-order methods. In addition, the partial connection in the bottom layer could be extended to higher layers as partial connectivities have many advantages, such as lower complexity and higher generalization ability.
Though this method can extract non-linear features, it is only very effective when dealing with advertisements without images. Consequently, further research on multi-modal sponsored search to model images and text would be useful to pursue.

\textbf{Summarizing Retrieved Documents}.
\label{section:nn-task-summarization}
While IR is typically understood to stop at retrieving relevant documents, {\bf \citet{lioma_deep_2016}} instead propose to synthesize retrieved documents, generating a new document for the user. They use an LSTM, trained using Torch\footnote{\url{https://github.com/jcjohnson/torch-rnn}}, to generate the synthetic documents. The LSTM takes the concatenated text of the query and its known relevant documents using word embeddings as input. However, rather than taking the whole text of a document, the authors extract a context window of $\pm n$ terms around every query term. 
To evaluate synthetic documents, the authors collect user evaluations through CloudFlower\footnote{\url{https://www.crowdflower.com}}, defining four queries per job. Each user is presented the query text and four wordclouds, generated from the synthetic document and three random relevant documents. Though deeper semantics are lost, wordclouds are selected for ease of the users, and each user is tasked with ordering the wordclouds from most to least relevant to the query. 
Experimental results suggest that the deep learned synthetic documents were on average considered to be more relevant than existing indexed relevant documents.
For future work, \citet{lioma_deep_2016} plan to experiment with word-level RNNs to produce better quality synthetic documents, as character-level RNNs have been criticized for producing noisy pseudo-terms.

\textbf{Temporal IR.}
Temporal IR is concerned with the temporal relevance of documents. For example, some queries are related to real events in the world and are expected to be matched to documents also pertaining to the same event more so than to static elements of the query. {\bf \citet{kanhabua_learning_2016}} use a deep NN to classify when a document is related to a temporal event and to identify the type of temporal event (anticipated, breaking, commemorative, meme, or ongoing). Hand-coded temporal query log features are used as input, such as burst length, average frequency, existence of person or location entities, auto-correlation, and click entropy. Despite manually-labeled target data being quite sparse, their deep network outperforms shallower baselines like Naive Bayes \cite{rish2001empirical}, Multi-layer Perceptron (MLP), and LibSVM \cite{chang2011libsvm}. 
However, it is unclear how hand-coded features would be obtained for dynamic events in order to detect them.

\section{Discussion} 
\label{section:methods}


\subsection{Word Embedding}
\label{section:method-embedding}

Use of word embeddings involves many design decisions. Should one use \wv{} (CBOW or \skipgram{}) \cite{mikolov2013distributed,mikolov2013efficient}, GloVe~\cite{pennington2014glove}, or something else (such as count-based embeddings)? Can embeddings from multiple sets be selected 
dynamically or combined together~\cite{neelakantan-EtAl:2014:EMNLP2014,zhang-roller-wallace:2016:N16-1}? How should hyper-parameters be set (e.g., dimensionality of embedding space, window size, etc.)?  What training data/corpora should be used?  Presumably larger training data is better, along with in-domain data similar to the test data on which the given system is to be applied, but how does one trade-off greater size of out-of-domain data vs.\ smaller in-domain data? How might they be best used in combination? Does performance vary much if we simply use off-the-shelf embeddings (e.g., from \wv{} or GLoVe) vs.\ re-training embeddings for a target domain, either by {\em fine-tuning}~\cite{mesnil2013investigation} off-the-shelf embeddings or re-training from scratch? How much does task (e.g., \adhoc{} search vs.\ document classification) or downstream architecture matter, e.g., will word embeddings be used directly in a traditional IR or machine learning model, or will they be input to an NN architecture for {\em end-to-end} learning? How should one deal with out-of-vocabulary (OOV) query terms not found in the word embedding training data for query-document matching?

\citet{vulic_monolingual_2015,yang_using_2016,zheng_learning_2015,zuccon_integrating_2015} compare different training hyper-parameters such as window size and dimensionality of word embeddings. \citet{zuccon_integrating_2015}'s sensitivity analysis of the various model hyperparameters for inducing word embeddings shows that manipulations of embedding dimensionality, context window size, and model objective (CBOW vs \skipgram{}) have no consistent impact on model performance vs.\ baselines. \citet{vulic_monolingual_2015} find that while increasing dimensionality provides more semantic expressiveness, the impact on retrieval performance is relatively small. \citet{zheng_learning_2015} find that 100 dimensions work best for estimating term weights, better than 300 and 500. In experiments using Terrier, \citet{yang_using_2016} find that for the Twitter election classification task using CNNs, word embeddings with a large context window and dimension size can achieve statistically significant improvements.

Selection among \wv{} CBOW or \skipgram{} or GloVe appears quite varied. \citet{zuccon_integrating_2015} compare CBOW vs.\ \skipgram{}, finding ``no statistical significant differences between the two...''  \citet{kenter_short_2015} use both \wv{} and GloVe \cite{pennington2014glove} embeddings (both originally released embeddings as well training their own embeddings) in order to induce features for their machine learning model. They report model effectiveness using the original public embeddings with or without their own additional embeddings, but do not report further ablation studies to understand the relative contribution of different embeddings used. \citet{grbovic_search_2015}'s \qv{} uses a two-level architecture in which the upper layer models the temporal context of query sequences via \skipgram{}, while the bottom layer models word sequences within a query using CBOW. However, these choices are not justified, and their later work \cite{grbovic_context-and_2015} uses \skipgram{} only. \citet{almasri_comparison_2016} evaluate both CBOW and \skipgram{} \wv{} embeddings (using default dimensionality and context window settings) but present only \skipgram{} results, writing that ``there was no big difference in retrieval performance between the two.'' \citet{zamani_estimating_2016,zamani_embedding-based_2016} adopt GloVe without explanation. Similarly for \wv{}, \citet{mitra_dual_2016} simply adopt CBOW, while  \citet{de_vine_medical_2014,manotumruksa_modelling_2016,vulic_monolingual_2015,yang_anmm:_2016,ye_word_2016,zhou_learning_2015} adopt \skipgram{}. \citet{zhang2015sensitivity} performed an empirical analysis in the context of using CNNs for short text classification. They found that the ``best'' embedding to use for initialization depended on the dataset. Motivated by this observation, the authors proposed a method for jointly exploiting multiple sets of embeddings (e.g., one embedding set induced using GloVe on some corpus and another using a \wv{} variant on a different corpus) \cite{zhang-roller-wallace:2016:N16-1}. This may also be fruitful for IR tasks, suggesting a potential direction for future work.

In selecting training data for word embeddings, \citet{yang_using_2016} considers classification in which one background corpus (used to train word embeddings) is a Spanish Wikipedia Dump which contains over 1 million articles, while another is a collection of 20 million tweets having more than 10 words per tweet. As expected, they find that when the background training text matches the classification text, the performance is improved. On the other hand, \citet{zuccon_integrating_2015} also consider different training corpora, but find that ``the choice of corpus used to construct word embeddings had little effect on retrieval results.'' \citet{zamani_estimating_2016} train GloVe on three external corpora and report, ``there is no significant differences between the values obtained by employing different corpora for learning the
embedding vectors.'' Regarding their model, \citet{zheng_learning_2015} write:
\begin{quotation}
\noindent ``[the system] performed equally well with all three external corpora; the differences among them were too small and inconsistent to support any conclusion about which is best. However, although no external corpus was best for all datasets... The corpus-specific word vectors were never best in these experiments... given the wide range of training data sizes -- varying from 250 million words to 100 billion words -- it is striking how little correlation there is between search accuracy and the amount of training data.''
\end{quotation}

%
Regarding handling of OOV terms, the easiest solution is to discard or ignore the OOV terms. For example, \citet{zamani_estimating_2016} only consider the queries where the embedding vectors of all terms are available. However, in end-to-end systems, where we are jointly estimating (or refining) embeddings alongside other model parameters, it is intuitive to randomly initialize embeddings for OOV words. For instance, in the context of CNNs for text classification, \citet{kim2014convolutional} adopted this approach. The intuition behind this is two-fold. First, if the same OOV appears in a pair of texts, queries, or documents being compared, this contributes to the similarity scores between those two. Second, if two different OOV terms appear in the same pair, their dissimilarity will not contribute in the similarity function. However, this does not specifically address accidental misspellings or social spellings (e.g., ``kool'') commonly found in social media. One might address this by hashing words to character n-grams (see Section~\ref{section:method-hashing}) or character-based modeling more generally (e.g., \citet{dhingra-EtAl:2016:P16-2,zhang2015character}, and \citet{conneau2016very}).

Particularly notable uses of embeddings include \citet{mitra_dual_2016}'s exploiting of both the input and the output projections. They map query terms into the input space and the document terms into the output space. The motivation for this is that they can thus use both the embedding spaces to better capture the relatedness between queries and documents. \citet{diaz_query_2016}'s work is unique in developing topic-specific embeddings specific to a query (by training off a weighted sampled of its retrieved documents). They incorporate the new embedding into the original language model to get a better query language model for query expansion. \citet{vulic_monolingual_2015} learn bilingual word embeddings from comparable corpora rather than parallel corpora via a merge-and-shuffle approach, applying their model to cross-lingual IR.

Several authors train embeddings on query logs rather than documents \cite{mitra_exploring_2015,mitra_query_2015,mitra_dual_2016,sordoni_learning_2014,sordoni_hierarchical_2015}. \citet{grbovic_context-and_2015} train embeddings using {\tt paragraph2vec} \cite{le_distributed_2014} on the query and their session to obtain better query embeddings. They also propose the addition of ad clicks and search result clicks to the query context. This can help with query rewriting, which in turn can help retrieve more relevant ads. 

Word embeddings have been learned from many other genres of text as well.
\citet{zhou_learning_2015} use the \wv{} \skipgram{} model to train word embeddings on community questions and their metadata to get better embeddings for question terms. They constrain the words from the same category to be close to each other so that the trained word embedding can encode categorical information. 
\citet{manotumruksa_modelling_2016} use \skipgram{} to train word embeddings on venues' comments dataset from Foursquare to get better representation for venues and users' preferences, which they later use for venue recommendation.
\citet{de_vine_medical_2014} use \skipgram{} to train on journal abstracts and patient records corpora to get embeddings specific for medical terms. 

Similar to past development of new modeling techniques in IR, there is a common theme of researchers starting first with bag-of-words models then wanting to move toward modeling longer phrases in their future work. \citet{ganguly_word_2015} suggest future work should investigate compositionality of term embeddings.  \citet{zuccon_integrating_2015} propose incorporating distributed representations of phrases to better model query term dependencies and compositionality. \citet{zheng_learning_2015} propose direct modeling of bigrams and proximity terms. \citet{zamani_estimating_2016} suggest query language models based on mutual-information and more complex language models (bigram, trigram, etc.) could be pursued. \citet{kenter_short_2015} note that future work could extend their approach to consider word order.     

In organizing this literature review, we have conceptually distinguished studies using word embeddings as inputs to an NN architecture (Section~\ref{section:nn}). In this case, the authors may further {\em fine-tune} word embeddings during NN training. For example, \citet{sordoni_hierarchical_2015} feeds embeddings into an RNN and then tunes embeddings during RNN training. On the other hand, \citet{yang_anmm:_2016} firstly {\em pre-train} word embeddings using the \wv{} \skipgram{} model as the input to their neural network, then fix them during training. However, there is not yet 
any clear evidence showing that either method consistently works best. 

\subsection{Word Hashing}
\label{section:method-hashing}

A number of papers \cite{shen_learning_2014,shen_latent_2014,ye_learning_2015,mitra_exploring_2015,mitra_query_2015} from Microsoft Research build on the DSSM model \cite{huang_learning_2013}, likely due in part to DSSM's authors kindly sharing their learned model for others to use. DSSM is additionally notable in using word hashing to convert words into character n-grams, which are used as the input to neural network. 
Words are broken down into letter n-grams and then represented as a vector of letter n-grams. \cite{huang_learning_2013} provide empirical analysis where the original vocabulary size of $500k$ is reduced to only $30k$ because of word hashing. While the number of English words can be unlimited, the number of letter n-grams in English is often limited, thus word hashing can resolve the out-of-vocabulary (OOV) problem as well (Section ~\ref{section:method-embedding}). However, one inherent problem of word hashing is the hashing conflict which can be serious for a very large corpus. 
We note that this technique precedes the advent of \citet{mikolov2013distributed,mikolov2013efficient})'s \wv{}, and most studies since then use word embeddings rather than word hashing to encode words. 

\subsection{Representing Text Beyond Words}
\label{section:method-representation}

A simple way to represent longer textual units, such as phrases, sentences, or entire documents, is to simply sum or average their constituent word embeddings. However such bag-of-words composition ignores word ordering, and simple averaging treats all words as equally important in composition (though some work has considered weighted operations, e.g., \citet{vulic_monolingual_2015}). \citet{ganguly_representing_2016} opine that:
\begin{quotation}
\noindent ``...adding the constituent word vectors... to obtain the vector  representation of the whole document is not likely to be useful, because... compositionality of the word vectors [only] works well when applied over a relatively small number of words... [and] does not scale well for a larger unit of text, such as passages or full documents, because of the broad context present within a whole document.''
\end{quotation}
Fortunately, a variety of alternative methods have been proposed for inducing representations of such longer textual units. However, there is no evidence that any one method performs consistently best, with the performance of each method instead appearing to often depend on the specific task and dataset being studied. We further discuss such methods below.


\citet{le_distributed_2014}'s oft-cited \pv{} (PV) trains a fixed-length vector representation for sentences and documents with two variants. The first proposed variant is Distributed Memory Model of PVs (PV-DM), which treats each paragraph as a unique token and concatenates/averages it with the word vectors in a context window in this paragraph. It then uses the concatenation/average to predict the following word in the context window. At test time, it needs to do inference on the unseen paragraph while fixing the word vectors already trained. The second proposed variant is Distributed Bag-of-Words (PV-DBOW), which uses PV to predict the word randomly sampled from the paragraph. Both variants train a fixed-length vector for sentences and documents, which can then be fed into a standard classifier like logistic regression. 
While \pv{} has been adopted in many studies (e.g., \citet{xu_modeling_2016} use PV-DBOW to represent documents) and is often reported as a baseline (e.g., \cite{tai-socher-manning:2015:ACL-IJCNLP}), concerns about reproducibility have also been raised. \citet{kiros2015skip} report results below SVM when re-implementing \pv{}. \citet{kenter_short_2015} note, ``it is not clear, algorithmically, how the second step -- the inference for new, unseen texts -- should be carried out.'' Perhaps most significantly, later work co-authored by Mikolov \cite{mesnil2014ensemble}\footnote{\url{https://github.com/mesnilgr/iclr15}} has disavowed the original findings of \citet{le_distributed_2014}, writing, ``to match the results from \citet{le_distributed_2014}, we followed the [author's] suggestion...  However, this produces the [reported] result only when the training and test data are not shuffled. Thus, we consider this result to be invalid.''

\citet{balikas_empirical_2016} propose large-scale text classification using distributed document-level representations. The document-level representations are obtained from \wv{} \skipgram{} word embeddings using naturally parallelizable composition functions based on neural networks. They further propose improving this method by combining document-level representations with traditional one-hot-encodings.
\citet{salakhutdinov_semantic_2009} propose an auto-encoder approach to model documents. \citet{gupta_query_2014} also use auto-encoding to jointly model terms across mixed-scripts.

\subsubsection{Convolutional Neural Networks (CNNs)}
\label{section:method-cnn}

\citet{shen_latent_2014,shen_learning_2014} propose a Convolutional Latent Semantic Model (CLSM) to encode queries and documents into fix-length vectors, following a popular ``convolution+pooling'' CNN architecture. The first layer of CLSM is a word hashing layer that can encode words into vectors. 
CLSM does not utilize word embeddings as input, seemingly distinguishing it from all other works using CNNs. 
\citet{mitra_exploring_2015} use CLSM to encode query reformulations for query prediction, while  
\citet{mitra_query_2015} use CLSM on query prefix-suffix pairs corpus for query auto-completion. They sample queries from the search query logs and split the query at every possible word boundary to form prefix-suffix pairs.  

\citet{severyn_learning_2015} and \citet{suggu_deep_2016} adopt a similar ``convolution+pooling'' CNN architecture to encode question and answer sentence representations, which serve as features for a second-stage ranking NN. See \citet{mitra_dual_2016} for further discussion of such two-stage {\em telescoping} approaches.  

\citet{yang_anmm:_2016} develop a novel value-shared CNN, and apply it on the query-answer matching matrix to extract the semantic matching between the query and answer. This model can capture the interaction between intermediate terms in the query and answer, rather than only considering the final representation of the query and answer. The motivation behind the value-shared CNN is that semantic matching value regularities
between a question and answer is more important than spatial regularities typical in computer vision. Similarly, contemporaneous work by \citet{guo_deep_2016} notes:
\begin{quotation}
\noindent ``Existing interaction-focused models, e.g., ARC-II and MatchPyramid, employ a CNN to learn hierarchical matching patterns over the matching matrix. These models are basically position-aware using convolutional units with a local ``receptive field'' and learning positional regularities in matching patterns. This may be suitable for the image recognition task, and work well on semantic matching problems due to the global matching requirement (i.e., all the positions are important). However, it may not be suitable for the ad-hoc retrieval task, since such positional regularity may not exist...''
\end{quotation}

\noindent \citet{cohen_adaptability_2016} compare CNNs vs.\ RNNs based on document length (see Section \ref{section:method-granularity}).

\subsubsection{Recurrent Neural Networks (RNNs)}
\label{section:method-rnn}

\citet{sordoni_hierarchical_2015} build a hierarchical GRU to encode the query and the query session into vector representations. \citet{song_multi-rate_2016} use an LSTM to model user interests at different time steps and encode them into a vector.
\citet{lioma_deep_2016} create new relevant information using an LSTM that takes the concatenated text of a query and its known relevant documents as input using word embeddings. However, rather than take the whole text of a document, they extract a context window of 
$\pm n$ terms around every query term occurrence. 
\citet{yan_learning_2016} use a bidirectional LSTM followed by a CNN to model the original query, reformulated query, candidate reply, and antecedent post in their human-computer conversation system.
\citet{wang2015long} use a stacked bidirectional LSTM to sequentially read words from both question and answer sentences, calculating relevance scores for answer sentence selection through mean pooling across all time steps. Section \ref{section:method-granularity} presents \citet{cohen_adaptability_2016}'s comparison of CNNs vs.\ RNNs by document length.

\subsubsection{Text Granularity in IR}
\label{section:method-granularity}

It is notable that many studies to date focus on short text matching rather than longer documents more typical of modern IR \adhoc{} search. \citet{cohen_adaptability_2016} study how the effectiveness of NN models for IR vary as a function of document length (i.e., text granularity). They consider three levels of granularity: i) \textit{fine}, where documents often contain only a single sentence and relevant passages span only a few words (e.g., question answering); ii) \textit{medium}, where documents consist of passages with a mean length of 75 characters and relevant information may span multiple sentences (e.g., passage retrieval); and iii) \textit{coarse}, or typical modern \adhoc{} retrieval. For fine granularity, they evaluate models using the TREC QA dataset and find that CNNs outperform RNNs and LSTM, as their filter lengths are able to effectively capture language dependencies. For medium granularity, they evaluate using the Yahoo Webscope L4 CQA dataset and conclude that LSTM networks outperform CNNs due to their ability to model syntactic and semantic dependencies independent of position in sequence. In contrast, RNNs without LSTM cells do not perform as well, as they tend to ``forget'' information due to passage length. 

For \adhoc{} retrieval performance on Robust04, comparing RNNs, CNNs, LSTM, DSSM, CLSM, \citet{vulic_monolingual_2015}'s approach, and \citet{le_distributed_2014}'s \pv{}, \citet{cohen_adaptability_2016} find that all neural models perform poorly. They note that neural methods often convert documents into fixed-length vectors, which can introduce a bias for either short or long documents. However, they find that the approaches of \citet{vulic_monolingual_2015} and \citet{le_distributed_2014} perform well when combined with language modeling approaches which explicitly capture matching information of queries and documents.

\subsection{Measuring Textual Similarity}


Given a text representation, perhaps using word embeddings, how do we measure textual similarity? Similarity here is not strictly limited to linguistic semantics, but more generally includes relevance matching between queries and documents or questions and answers. While cosine similarity is the simplest and most common approach,
a variety of more sophisticated methods have been proposed.

As noted at the start of Section~\ref{section:method-representation}, \citet{ganguly_representing_2016} opine that simply adding vectors of word embedding cannot sufficiently capture context of longer textual units. Instead, the authors propose a new form of similarity metric based on the assumption that the document can be represented as a mixture of p-dimensional Gaussian probability density functions and each \wv{} embedding (p-dimensions) is an observed sample. Then, using the EM algorithm, they estimate the probability density function which can be incorporated to the query likelihood language model using linear interpolation. 

\citet{zamani_embedding-based_2016} and \cite{zamani_estimating_2016} propose using sigmoid and softmax transformations of cosine similarity on the grounds that cosine similarity values are not discriminative enough \cite{zamani_embedding-based_2016}. Their empirical analysis shows that there are no substantial differences (e.g., two
times more) between the similarity of the most similar term
and the $1000$th similar term to a given term $w$, while the
$1000$th word is unlikely to have any semantic similarity with
$w$. Consequently, they propose using monotone mapping functions (e.g., sigmoid or softmax) to transform the cosine similarity scores.

\citet{kenter_short_2015} propose a BM25 extension to incorporate word embeddings in order to calculate the similarity of short text. 
Their approach uses a word alignment method and a saliency-weighted semantic graph to 
move from word-level to text-level semantics. Features are computed from the word alignment method and from the means of word embeddings to train a final classifier that predicts a semantic similarity score.

\citet{kusner_word_2015} develop word movers distance (WMD) for computing distance between documents. It is observed by \citet{mikolov2013distributed} that semantic relationships are often preserved in vector operations on \wv{}. Thus, the proposed method \cite{kusner_word_2015} utilizes this property of \wv{} (see Section \ref{section:we-background}) and computes the minimum traveling distance from the embedded words of one document to another. This process is modeled as an Earth Mover's Distance (EMD) problem, a well studied optimization problem. 
\citet{kim_bridging_2016} propose another version of WMD specifically for query-document similarity measure. They handle the high computational cost of WMD by mapping queries to documents using a word embedding model trained on a document set. They make several changes to the original WMD methods: changing the weight of term by introducing inverse document frequency, and changing the original dissimilarity measure to cosine similarity. However, they do not provide any comparison vs.\ WMD as a baseline.  

\citet{zhang_continuous_2014} use Fisher Kernel \cite{jaakkola1999exploiting} to calculate the similarity between short text segments given a trained word embedding to detect local text reuse. \citet{clinchant_aggregating_2013} and \citet{zhou_learning_2015} also adopt a Fisher Kernel approach.  
\citet{rekabsaz_uncertainty_2016} 
develop a global similarity threshold to filter highly related terms having an expected number of synonyms for a word.
\citet{sordoni_learning_2014} embed queries and documents in a larger space than single terms on the grounds that text sequences have more informative content. Documents are scored for retrieval using a {\em quantum relative entropy} approach. 
\citet{wang2015long} use a stacked bidirectional LSTM to sequentially read words from both question and answer sentences, calculating relevance scores for answer sentence selection through mean pooling across all time steps. 
\citet{yan_learning_2016} match sentences by concatenating their vector representations and feeding them into a multi-layer fully-connected neural network, matching a query with the posting and reply in a human computer conversation system.
\citet{xu_modeling_2016} propose a neural tensor network (NTN) approach to model document novelty. This model takes a document and a set of other documents as input. The architecture uses a tensor layer, a max-pooling layer, and a linear layer to output a document novelty score. 

\citet{guo_deep_2016}'s recent Deep Relevance Matching Model (DRMM) appears to be one of the most successful to date for \adhoc{} search over longer document lengths. In terms of textual similarity, they argue that the \adhoc{} retrieval task is mainly about relevance matching, different from semantic matching in NLP. They model the interaction between query terms and document terms, building a matching histogram on top of the similarities. They then feed the histogram into a feed forward neural network. They also use a term gating network to model the importance of each query term. 

\citet{yang_anmm:_2016} propose an attention-based neural matching model (aNMM) for question answering. Similar to \citet{guo_deep_2016}, they first model the interaction between query terms and document terms to build a matching matrix. They then apply a novel value-shared CNN on the matrix. Since not all the query terms are equally important, they use a softmax gate function as an attention mechanism in order to learn the importance of each query term when calculating the matching between the query and the answer.

\begin{longtable}{| l | p{9.0cm} |}
    \hline
    {\bf English Data} & \textbf{Study}  \\ \hline
	20-Newsgroup Corpus & \citet{salakhutdinov_semantic_2009} \\ \hline
    AOL Query Logs & \citet{kanhabua_learning_2016}, \citet{mitra_exploring_2015},  \citet{mitra_query_2015}, \citet{sordoni_hierarchical_2015} \\ \hline
    Bing Query Logs and WebCrawl & \citet{mitra_exploring_2015},  \citet{mitra_query_2015}, \citet{mitra_dual_2016}, \citet{nalisnick_improving_2016} \\ \hline
    CLEF03 English \Adhoc{} & \citet{clinchant_aggregating_2013} \\ \hline
    CLEF Medical Corpora & \citet{almasri_comparison_2016} \\ \hline
    CLEF 2016 Social Book Search  & \citet{amer_toward_2016} \\ \hline
    MSR Paraphrase Corpus & \citet{kenter_short_2015}\\ \hline
    MSN Query Log & \citet{kanhabua_learning_2016} \\ \hline
    OHSUMED & \citet{de_vine_medical_2014} \\ \hline
    PubMed & \citet{balikas_empirical_2016} \\ \hline
    Reuters Volume I (RCV1-v2) & \citet{salakhutdinov_semantic_2009} \\ \hline
    SemEval 2015-2016& DFFN~(\citet{suggu_deep_2016}) \\ \hline
    TIPSTER (Volume 1-3)& \citet{zhang_continuous_2014} \\ \hline
    TREC 1-2 \Adhoc{} (AP 88-89)& \citet{clinchant_aggregating_2013}, \citet{zamani_embedding-based_2016}, \citet{zamani_estimating_2016}, \citet{zuccon_integrating_2015} \\ \hline
    TREC 1-3 \Adhoc{} &  \citet{zuccon_integrating_2015} \\ \hline
     TREC 6-8 \Adhoc{} & GLM~(\citet{ganguly_word_2015}), \citet{lioma_deep_2016}, \citet{rekabsaz_uncertainty_2016}, \citet{roy_using_2016}, \\ \hline
     TREC 12 \Adhoc{} 	& \citet{diaz_query_2016} \\ \hline
     TREC 2005 HARD  & \citet{rekabsaz_uncertainty_2016} \\ \hline 
     TREC 2007-2008 Million Query& \citet{yang_selective_2016} \\ \hline
     TREC 2009-2011 Web & \citet{xu_modeling_2016} \\ \hline
     TREC 2009-2013 Web  & \pv{}~(\citet{grbovic_context-and_2015}) \\ \hline
     TREC 2010-2012 Web & QEM (\citet{sordoni_learning_2014}) \\ \hline
     TREC 2011 Microblog & CDNN~(\citet{severyn_learning_2015}),  \citet{zhang_continuous_2014} \\ \hline
     TREC 2012 Microblog & CDNN~(\citet{severyn_learning_2015}) \\ \hline
    TREC 2015 Contextual Suggestion & \citet{manotumruksa_modelling_2016} \\ \hline
    TREC \clueweb{} & DRMM~(\citet{guo_deep_2016}),  QEM (\citet{sordoni_learning_2014}), \citet{zhang_continuous_2014}, \citet{zheng_learning_2015}\\ \hline
    TREC DOTGOV & \citet{zuccon_integrating_2015} \\ \hline
    TREC GOV2 & \citet{yang_selective_2016}, \citet{zamani_embedding-based_2016}, \citet{zamani_estimating_2016}, \citet{zheng_learning_2015} \\ \hline
    TREC QA 8-13
& aNMM~(\citet{yang_anmm:_2016}), BLSTM~(\citet{wang2015long}), CDNN~(\citet{severyn_learning_2015}),   \citet{yu2014deep}, \cite{cohen_adaptability_2016} \\ \hline
    TREC MedTrack & \citet{de_vine_medical_2014}, \citet{zuccon_integrating_2015} \\ \hline
    TREC Robust & GLM~(\citet{ganguly_word_2015}), \citet{roy_using_2016} \\ \hline
    TREC Robust 2004 & \citet{clinchant_aggregating_2013}, \citet{diaz_query_2016}, DRMM~(\citet{guo_deep_2016}), \citet{zamani_embedding-based_2016}, \citet{zamani_estimating_2016}, \citet{zheng_learning_2015} \\ \hline
    TREC WSJ87-92 & \citet{zuccon_integrating_2015} \\ \hline
    TREC WT10G &  \citet{roy_using_2016}, \citet{zheng_learning_2015}\\ \hline
    Yahoo! Answers & \citet{zhou_learning_2015}\\ \hline
    \multicolumn{2}{c}{} \\ \hline
	\textbf{Chinese Data} & {\bf Study}\\ \hline
	Baidu Tieba &  \citet{yan_learning_2016} \\ \hline
    Baidu Zhidao &  \citet{yan_learning_2016}, \citet{zhang_continuous_2014}, \citet{zhou_learning_2015} \\ \hline
    Douban Forum &  \citet{yan_learning_2016} \\ \hline
    Sina Weibo &  \citet{yan_learning_2016}, \citet{zhang_continuous_2014} \\ \hline
    SogouT 2.0 & \citet{zhang_continuous_2014} \\ \hline
    \multicolumn{2}{c}{} \\ \hline
	\textbf{Multi-Lingual Data} & {\bf Study}\\ \hline
    CLEF 2001-2003 \Adhoc{} &\citet{vulic_monolingual_2015} \\ \hline
    FIRE 2013 & \citet{gupta_query_2014} \\ \hline
    iPinYou \cite{liao2014ipinyou} &  \citet{zhang_deep_2016} \\ \hline
\caption{Datasets Used}
\label{table:Data_Used}
\end{longtable}


{
\renewcommand{\arraystretch}{1.3}
\begin{table}[ht]
\begin{center}
  \begin{tabular}{ | l | p{3.6cm} | p{8.5cm} | }
    \hline    
    \textbf{System}  & \textbf{Citation} & \textbf{URL} \\ \hline
    \wv{} & \citet{mikolov2013distributed} & \url{https://code.google.com/archive/p/word2vec/} \\
    GloVe & \citet{pennington2014glove} & \url{http://nlp.stanford.edu/projects/glove/} \\
    \hline
    CDNN & \citet{severyn_learning_2015}& \url{https://github.com/aseveryn/deep-qa} \\ 
    DeepMerge & \citet{lee_optimization_2015} & \url{https://ciir.cs.umass.edu/downloads/DeepMerge/} \\
    DeepTR & \citet{zheng_learning_2015} & \url{http://www.cs.cmu.edu/~gzheng/code/TermRecallKit-v2.tar.bz2} \\     
    Mixed Deep & \citet{gupta_query_2014} & \url{http://www.dsic.upv.es/~pgupta/mixed-script-ir} \\
    NTLM & \citet{zuccon_integrating_2015} & \url{https://github.com/ielab/adcs2015-NTLM} \\
    \hline
  \end{tabular}
\end{center}
\caption{Source Code Released}
\label{table:code}
\end{table}
}

{
\renewcommand{\arraystretch}{1.3}
\begin{table}[ht]
\begin{center}
  \begin{tabular}{ | l | p{3.0cm} | p{7.0cm} | }
    \hline
    \textbf{Data}  & \textbf{Citation} & \textbf{URL} \\ \hline
      \wv{} embeddings & \citet{mikolov2013distributed} & \url{https://code.google.com/archive/p/word2vec/} \\
      GloVe embeddings & \citet{pennington2014glove} & \url{http://nlp.stanford.edu/projects/glove/} \\
      \hline
      Bing query embeddings & \citet{mitra_dual_2016} & \url{https://www.microsoft.com/en-us/download/details.aspx?id=52597}\\ 
    NTLM embeddings & \citet{zuccon_integrating_2015} & \url{http://www.zuccon.net/ntlm.html} \\
    \hline
  \end{tabular}
\end{center}
\caption{Datasets Released}
\label{table:data}
\end{table}
}

\section{Conclusion}
\label{section:conclusion}

Interest in Neural IR has never been greater, spanning both active research and deployment in practice\footnote{\url{https://en.wikipedia.org/wiki/RankBrain}}~\cite{metz2016}. Neural IR continues to accelerate in quantity of work, sophistication of methods, and practical effectiveness (see \citet{guo_deep_2016}). New methods are being explored that may be computationally infeasible today (see \citet{diaz_query_2016}), but if proven effective, could motivate future optimization work to make them more practically viable (e.g., \cite{jurgovsky_evaluating_2016,ordentlich_network-efficient_2016}). In his opening Keynote at SIGIR 2016, Chris Manning noted the rise of NN approaches to dominance in speech recognition (2011), computer vision (2013), and NLP (2015), and concluded with an unabashed assertion that, ``I'm certain that deep learning will come to dominate SIGIR over the next couple of years... just like speech, vision, and NLP before it''~\cite{manning_sigir2016}.

At the same time, healthy skepticism about Neural IR also remains. Despite his bold assertion, Manning nevertheless reminded his audience to always be wary of the common ``emerging technology hype cycle'' of having unrealistic expectations early in the development of any promising new technology. The key question in IR today might be most succinctly expressed as: ``Will it work?''. While NN methods have worked quite well on short texts, effectiveness on longer texts typical of \adhoc{} search has been problematic \cite{huang_learning_2013,cohen_adaptability_2016}, with only very recent evidence to the contrary \cite{guo_deep_2016}. In addition, while great strides have been made in computer vision through employing a very large number of {\em hidden layers} (hence ``deep'' learning), such deep structures have typically been less effective in NLP and IR than shallower architectures~\cite{pang_study_2016}, though again with notable recent exceptions (see \cite{conneau2016very}). When Neural IR has improved in \adhoc{} search results, improvements appear relatively modest~\cite{zamani_embedding-based_2016,diaz_query_2016} when compared to traditional query expansion techniques for addressing {\em vocabulary mismatch}, such as pseudo-relevance feedback (PRF). Both \citet{ganguly_representing_2016} and \citet{diaz_query_2016} have noted that {\em global} word embeddings, trained without reference to user queries, vs.\ {\em local} methods like PRF for exploiting query-context, appear limited similarly to the traditional global-local divide seen with existing approaches like topic modeling~\cite{yi2009comparative}.

As \citet{li_does_2016} so eloquently put it, ``Does IR Need Deep Learning?'' Such a seemingly simple question requires careful unpacking. Much of the above discussion assumes Neural IR should deliver new state-of-the-art quality of search results for traditional search tasks. While it may do so, this framing may be far too narrow, as \citet{li_does_2016}'s presentation suggests. The great strength of Neural IR may lie in enabling a new generation of search scenarios and modalities, such as searching via conversational agents~\cite{yan_learning_2016}, multi-modal retrieval~\cite{ma_multimodal_2015,ma_learning_2015}, knowledge-based search IR~\cite{nguyen_toward_2016}, or synthesis of relevant material~\cite{lioma_deep_2016}. It may also be that Neural IR will provide greater traction for other future search scenarios not yet considered.  

Given that efficacy of deep learning approaches is often driven by ``big data'', will Neural IR represent yet another fork in the road between industry and academic research, where massive commercial query logs deliver Neural IR's true potential? There is also an important contrast to note here between supervised scenarios, such as learning to rank~\cite{liu2009learning} vs.\ unsupervised learning of word embeddings or typical queries (see \citet{mitra_query_2015,mitra_exploring_2015,sordoni_hierarchical_2015}). \citet{lecun_deep_2015} wrote, ``we expect unsupervised learning to become far more important in the longer term.'' Just as the rise of the Web drove work on unsupervised and semi-supervised approaches by the sheer volume of unlabeled data it made available, the greatest value of Neural IR may naturally arise where the biggest data is found: continually generated and ever-growing behavioral traces in search logs, as well as ever-growing online content. 

While skepticism of Neural IR may well remain for some time, the practical importance of search today, coupled with the potential for significantly new traction offered by this ``third wave'' of NNs, makes it unlikely that researchers will abandon Neural IR anytime soon without having first exhaustively tested its limits. As such, we expect the pace and interest in Neural IR will only continue to blossom, both in new research and increasing application in practice. 


\section{Additional Authors}
\label{section:authors}

The following additional students at the University of Texas at Austin contributed indirectly to the writing of this literature review: Manu Agarwal,
Edward Babbe,
Anuparna Banerjee,
Jason Cai,
Dillon Caryl,
Yung-sheng Chang,
Shobhit Chaurasia,
Linli Ding,
Brian Eggert,
Michael Feilbach,
Alan Gee,
Jeremy Gin,
Rahul Huilgol,
Miles Hutson,
Neha Javalagi,
Yan Jiang,
Kunal Lad,
Yang Liu,
Amanda Lucio,
Kristen Moor,
Daniel Nelson,
Geoffrey Potter,
Harshal Priyadarshi,
Vedhapriya Raman,
Eric Roquemore,
Juliette Seive,
Abhishek Sinha,
Ashwini Venkatesh,
Yuxuan Wang,
and Xu Zhang.
%





\bibliography{zotero,notinzotero}  
\bibliographystyle{apa}


\end{document}